\documentclass[a4paper,12pt]{article}
\usepackage{amsmath,amsthm,amsfonts,amssymb,bm,mathrsfs}
\usepackage{mathtools}
\usepackage{graphicx}
\usepackage{xcolor}
\usepackage{float}
\usepackage[super,square,numbers,sort&compress]{natbib} 
\usepackage{fancyhdr}
\usepackage{chemarrow}
\usepackage{extarrows}
\usepackage{enumerate}
\usepackage{wrapfig}
\usepackage{bibentry,natbib}
\setcitestyle{authoryear,open={(},close={)}}
\usepackage[margin=0.5in]{geometry}
\usepackage[T1]{fontenc}
\usepackage{footmisc}
\usepackage{sgame}
\usepackage{color}
\usepackage{multicol}
\usepackage{tikz}
\usetikzlibrary{decorations.pathreplacing}
\usetikzlibrary{arrows}
\tikzstyle{block}=[draw opacity=0.7,line width=1.4cm]
\usepackage{pgfplots}
\pgfplotsset{width=10cm,compat=1.9}
\usepgfplotslibrary{external}
\usepackage{fancyref}
\usepackage[colorlinks=true, linkcolor=blue, urlcolor=blue, citecolor=blue]%
{hyperref}
\usepackage{amsmath}
\usepackage{caption}
\usepackage{amsfonts}
\usepackage{amssymb}
\usepackage{lmodern}
\usepackage{enumitem}
\usepackage{dsfont}
\providecommand{\U}[1]{\protect\rule{.1in}{.1in}}
\allowdisplaybreaks
\usetikzlibrary{matrix,arrows,decorations.pathmorphing}

\newtheorem{theorem}{Theorem}
\newtheorem{assumption}{Assumption}

\newtheorem{definition}{Definition}

\newtheorem{lemma}{Lemma}

\newtheorem{proposition}{Proposition}

\begin{document}
\raggedbottom
\title{\Large \textbf{A Theory of Updating Ambiguous Information}\thanks{I am deeply indebted to Faruk Gul, Pietro Ortoleva and Wolfgang Pesendorfer for their invaluable advice, guidance and encouragement. I thank Roland B\'enabou, Sylvain Chassang, Xiaoyu Cheng, Xiaosheng Mu, John K.-H. Quah, Satoru Takahashi, Mu Zhang and seminar participants at Princeton Microeconomic Theory Student Lunch Seminar for helpful comments and discussions. All errors are my own.}}

\author{
Rui Tang \thanks{Department of Economics, Princeton University, ruit@princeton.edu}
}

\maketitle

\begin{abstract}
We introduce a new updating rule, the conditional maximum likelihood rule (CML) for updating ambiguous information. The CML formula replaces the likelihood term in Bayes' rule with the maximal likelihood of the given  signal conditional on the state. We show that CML satisfies a new axiom, \text{increased sensitivity after updating}, while other updating rules do not. With CML, a decision maker's  posterior is unaffected by the order in which independent signals arrive. CML also accommodates recent experimental findings on updating signals of unknown accuracy and has simple predictions on learning with such signals. We show that an information designer can almost achieve her maximal payoff with a suitable ambiguous information structure whenever the agent updates according to CML. 
\bigskip

\textit{Keywords}: ambiguous information; conditional maximum likelihood;  non-dilation; increased sensitivity after updating; under-reaction to ambiguous information
\bigskip

\textit{JEL Codes}:  D01, D81, D93
\end{abstract}

\newpage
\baselineskip=24pt

\section{Introduction}\label{sec_intro}
In decision theory, the term ambiguity refers either to an event with unknown probability or to information that has multiple probabilistic interpretations. Since the seminal work of \citet*{qje1961ellsberg}, a variety of models have been proposed to rationalize decision makers' (henceforth DM) choices over bets on ambiguous events.\footnote{See, for instance, \citet*{jme1989maxmin}, \citet*{ecta1989choquet}, \citet*{ecta2006variational}, \citet*{chew2008small},  \citet*{ecta2014euu},  etc.}  In contrast, relatively few papers focus on updating ambiguous information. The growing literature on mechanism and information design under ambiguity has highlighted the importance of the latter topic. \medskip

In many applications (e.g., \citeauthor{ecta2014commu} \textcolor{blue}{2014} and \citeauthor*{jet2019pers} \textcolor{blue}{2019}), the DM's prior over the payoff-relevant state space is unambiguous but the information he receives is ambiguous. Our focus in this paper is on such applications. As in most existing models of choice under uncertainty, our DM is a max-min expected utility maximizer. That is, if the DM has a set of priors, he evaluates each ambiguous prospect according to its minimal expected utility over all possible  priors. In this paper, we offer an alternative to full-Bayesian updating (henceforth FB) where the set of posteriors is the set of Bayes' updates of the priors.\medskip

One consequence of FB is that a DM's set of posteriors may be a superset of his set of priors.  This {\it dilation} of beliefs may occur even when the DM has a single prior; that is, even if there is no payoff-relevant ambiguity.\footnote{For more discussions of dilation, see \citet*{as1990dilation}.}  \citet*{wp2019dilation} (henceforth SO19)  offer experimental evidence indicating that ambiguity averse DMs do not dilate after receiving ambiguous information. This finding is inconsistent with FB and other well-known updating rules such as the maximum likelihood rule (henceforth ML). Our new updating rule, CML, does not create dilation and is consistent with experimental evidence presented in recent papers.\medskip

To see how our updating rule works, consider the following example: there is a coin and an urn. The urn contains 100 balls; each ball is either red or blue. It is not known how many balls there are of either color. The coin is tossed, and a ball is drawn from the urn. Consider a bet that yields $\$20$ to the DM if the coin shows tails and $\$0$ if the coin shows heads. Before taking the bet, the DM receives a message about the outcome of the toss. The message is either ``tails'' or ``heads''. The message matches the outcome of the coin toss if a red ball is drawn and does not match the outcome if a blue ball is drawn. After receiving the message, the DM is asked if he would be willing to give up the bet in exchange for a objective lottery that yields $\$20$ with probability $1/2$ and $\$0$ with probability $1/2$. \medskip

Suppose that the DM receives the signal ``tails''. Given the signal, the bet depends on the color of the ball that is drawn: if the ball is red, the DM has won; if it is blue,  he has lost. Hence, it could be argued that once the signal is observed, the bet becomes ambiguous and therefore, an ambiguity averse DM might strictly prefer the objective lottery to the bet.  Full-Bayesian updating is consistent with this view:  with FB, an ambiguity averse DM would strictly prefer the objective lottery to the bet after receiving either signal ``tails'' or signal ``heads''.\medskip

Since the bet is equivalent to the objective lottery before the signal, the above argument suggests that ambiguity averse DMs lower their value of the bet after observing the signal. SO19, however, find that ambiguity averse DMs typically do not change their value of the bet after observing the signal. Our model provides the following rationale for the behavior of such DMs: if the outcome of the coin toss is tails,  then signal ``tails'' is most likely when the number of red balls in the urn is maximal. Symmetrically, if the outcome of the coin toss is heads, then the signal ``tails'' is most likely when the number of blue balls in the urn is maximal. Therefore, updating based on the conditionally most likely event, i.e., CML, preserves the initial symmetry and implies that the DM is indifferent between the bet and the objective lottery even after observing the signal.\medskip

Our model has a fundamental state space $S$ and a signal space $\Theta$.  We call $S\times \Theta$ the extended state space. A function mapping from the extended state space to payoffs is an extended act.  A function mapping from the  state space to payoffs is an act.  An extended evaluation function, $V$, describes the DM's ex-ante preferences over extended acts. An evaluation function, $V_{\theta}$,  describes the DM's ex-post preferences over acts after observing signal $\theta$.  Section \ref{sec_framework} provides the necessary formalism and an axiomatic characterization of max-min (extended) evaluation functions.\medskip

An updating rule specifies for each extended evaluation function $V$ and each signal $\theta$ an evaluation function $V_{\theta}$. CML is defined as follows. Consider an extended evaluation function $V$, which admits a max-min representation by a set of priors $\mathcal{P}$ over $S\times\Theta$.\footnote{That is, for any extended act $f^*$, $V(f^*)=\inf\limits_{p \in \mathcal{P}} \left(  \sum\limits_{(s,\theta) \in S\times\Theta} p(s,\theta) f^*(s,\theta)\right)$.} Throughout the paper, we assume that $\mathcal{P}$ has no ambiguity over $S$, i.e., the DM has a single prior over the state space.\footnote{$\mathcal{P}$ has no ambiguity over $S$ if for any $p,p' \in \mathcal{P}$, $p(\{s\}\times\Theta)=p'(\{s\}\times\Theta)$ for all $s \in S$.} For any signal $\theta$, the evaluation function $V_{\theta}$ specified by CML admits an expected utility representation by the posterior $\mu_{\theta}$ over $S$, where $\mu_\theta$ satisfies that $$\mu_{\theta}(s)=\frac{\max_{p \in \mathcal{P}} p(s,\theta)}{\sum_{s' \in S}\max_{p \in \mathcal{P}} p(s',\theta)}, \forall s \in S.\footnote{If $V_\theta$ admits an expected utility representation by $\mu_\theta$, then for each act $f$, $V_\theta(f)=\sum\limits_{s \in S} \mu_\theta(s) f(s)$.}$$
That is, following CML, a DM who has prior set $\mathcal{P}$ updates his belief over $S$ to $\mu_{\theta}$ after observing $\theta$. The maximization in the formula captures the conditional maximum likelihood of the signal on each state. \medskip

In Section \ref{sec_model}, we introduce a new axiom, {\it increased sensitivity after updating} (henceforth ISU), which distinguishes CML from existing updating rules. The idea of the axiom can be illustrated through the following example. Consider a state space containing three states $\{s_1, s_2, s_3\}$ with prior $(1/3, 1/3, 1/3)$. Let event $\{s_1, s_2\}$ be realized. The Bayes' posterior is given by $(1/2, 1/2)$.  Assume that the DM is an expected utility maximizer. If his payoff at state $s_1$ increases by $\epsilon > 0$, his ex-ante expected payoff increases by $1/3  \epsilon $, and his ex-post expected payoff increases by $1/2  \epsilon$. Obviously, he is more sensitive to payoff changes on $s_1$ after  event $\{s_1, s_2\}$ is realized. Axiom ISU is motivated by this observation. \medskip

We state axiom ISU in our framework. Consider two extended acts $f^*$ and $g^*$. Suppose that $g^*(s,\theta)> f^*(s,\theta)$ for some $(s,\theta) \in S\times\Theta$, and $g^*(s',\theta')=f^*(s',\theta')$ for all $(s',\theta')$ different from $(s,\theta)$. Let $f$ and $g$ be two acts satisfying $f(s)=f^*(\theta,s)$ and $g(s)=g^*(\theta,s)$ for each $s \in S$. An updating rule satisfies ISU if for any extended evaluation function $V$ and any signal $\theta$, 
\begin{equation*}\label{eq_intro_isu}
V(f^*)=V_{\theta}(f) \text{ implies } V(g^*)\le V_{\theta}(g),
\end{equation*}
where $V_{\theta}$ is the evaluation function specified by the updating rule. We interpret this axiom as follows. If the extended act changes from $f^*$ to $g^*$, the DM has a payoff increase in $(s, \theta)$. The payoff increase should affect the DM more after $\theta$ is observed, since it rules out event $S\times (\Theta\backslash \{\theta\})$ and increases the chance that $(s,\theta)$ occurs.  Therefore, the difference between the evaluations of $f^*$ and $g^*$ at the ex-post stage should be higher than that at the ex-ante stage. Since $f$ has the same evaluation as $f^*$, $g$ must have a higher evaluation than $g^*$. The condition $V(f^*)=V_{\theta}(f)$  further rules out the possibility that different sensitivities to payoff changes are caused by different ex-ante and ex-post utility levels. Axiom ISU is satisfied by CML but violated by other updating rules.   \medskip

In Section \ref{sec_characterization}, we characterize CML. We show that an updating rule is CML if and only if it satisfies axiom ISU, axiom independence of irrelevant signals, and axiom ratio consistency (Theorem \ref{thm_main}). The latter two axioms are also satisfied by FB. When the DM's extended evaluation function is not observed, we provide a characterization of CML rationalizable belief profiles. Specifically, a belief profile contains the DM's ex-ante belief over the state space and ex-post belief after observing each signal. We provide a sufficient and necessary condition for the existence of an information structure under which the belief profile is consistent with CML updating.\medskip

In Section \ref{sec_application}, we consider three applications of CML. First, we show that CML is divisible. That is, the DM's posterior is not affected by the order in which independent signals arrive. This makes CML suitable for the analysis of Wald-type problems under ambiguity. Second, we relate CML to experimental evidence involving signals of unknown accuracy. We show that a DM who updates according to CML  under-reacts to ambiguous information, as observed by \citet*{wp2019ambiguous} (henceforth L19).\footnote{See Section \ref{subsec_accuracy} for the definition of under-reaction to ambiguous information.} We also investigate the learning behavior of a DM who updates signals of unknown accuracy with CML. \medskip

Finally, we apply CML to a \citet*{aer2011persuasion}-style information design problem. We consider a designer who wants to persuade an agent to take an action that affects the designer's payoff. The designer can choose any ambiguous or unambiguous information structure.  We show that the designer can almost achieve her maximal payoff when facing a CML agent. That is, for any $\epsilon >0$, there exists an ambiguous information structure such that the designer's payoff is no less than her maximal payoff minus $\epsilon.$ \medskip

Section \ref{sec_discussion} contains our comparisons of CML with existing alternatives. We show that within the max-min expected utility framework, no updating rule satisfies axiom ISU if we allow multiple priors over the payoff-relevant state space.  Hence, we show that the ISU assumption is appropriate only when ambiguity arises from information. \bigskip

\textbf{Related Literature.}
A couple of experimental works directly test how subjects react to ambiguous infomation, e.g., \citet*{wp2019hard}, \citet*{wp2019ambiguous2}, etc. Our study is mostly related to the two experimental papers of SO19 and L19. In both SO19 and L19, the state space is finite. Both papers test how DMs react to ambiguous information when there is no ambiguity over the state space. Our theory can partially accommodate their findings. \medskip

Another stream of literature studies ambiguous information in game-theoretical frameworks, including \citet*{er2014vague}, \citet*{ecta2014commu}, \citet*{geb2017commu}, \citet*{jet2019pers}, \citet*{jet2018cheap}, etc. In this paper, we also apply CML to study the information design problem. We characterize the set of  payoffs that can be achieved by an information designer when the agent updates according to CML. \medskip

Our paper contributes to the literature of updating under ambiguity. The two most studied updating rules under ambiguity are FB and ML. FB is analyzed by \citet*{as1990dilation} and \citet*{ieee1992fb} and axiomatized by \citet*{td2002fb}. ML is introduced by \citet*{ams1967ml} and \citet*{book1976ml} and axiomatized by \citet*{jet1993ml} and \citet*{wp2019rml}. \citet*{wp2019rml} also provides a new updating rule, the relative maximum likelihood rule,  under which a DM's posterior set is a convex combination of FB and ML posterior sets. CML differs from these rules in two respects. First, CML satisfies axiom ISU. Second, with CML, the DM's prior over the state space is always updated to a single posterior and thus is not dilated. \medskip

\citet*{te2007dc} introduce the dynamic consistent updating rule, featuring a DM  who only updates a subset of his priors using Bayes' rule to maintain the optimality of the ex-ante optimal choice. The posterior set of a DM who updates according to \citet*{te2007dc}  depends on the ex-ante act as well as the choice menu, which violates consequentialism.\footnote{Consequentialism says that the ex-post evaluation of a choice does not depend on its payoffs on states not contained in the realized event.} CML differs from this rule in two respects. First, CML satisfies consequentialism. Second, the CML posterior of a DM may not be contained in his FB posterior set (Example 1).

A new updating rule, named the proxy rule, is proposed and axiomatized by \citet*{wp2018proxy}.  \citet*{wp2018proxy} introduce axiom ``not all news is bad news'' and show that the axiom is satisfied by the proxy rule but not FB or ML. A  common feature of the proxy rule and CML is that information does not render unambiguous events ambiguous. The key difference between CML and the proxy rule is that CML does not satisfy ``not all news is bad news'' while the proxy rule does not satisfy ISU.\footnote{See Section \ref{subsec_proxy} for a detailed discussion.} \medskip

The rest of the paper is organized as follows.  We present the framework of the paper in Section \ref{sec_framework}. We introduce CML and axiom ISU in Section \ref{sec_model}. We characterize CML in Section \ref{sec_characterization}. Section \ref{sec_application} contains applications of CML. We discuss CML in Section \ref{sec_discussion} and conclude the paper in Section \ref{sec_conclusion}. All omitted proofs are in the Appendix.

\section{Framework}\label{sec_framework}
\subsection{Preliminary}
\label{subsec_preliminary}
We define notations used in the paper. For an arbitrarily nonempty set $H$, let $\Delta(H)$ denote the set of finitely supported probability distributions over $H$, i.e., $d \in \Delta(H)$ if and only if there exists a nonempty and finite subset $H' \subseteq H$ such that $d(H')=1.$ If $H$ is finite, let $\Delta^o(H)$ be the relative interior of $\Delta(H)$, i.e., $d \in \Delta^o(H)$ if and only if $d(H') > 0$ for each nonempty $H' \subseteq H$. Any probability distribution over  $H' \subseteq H$ is considered as a probability distribution over $H$ that has support $H'$, and any probability distribution over $H$ that has support $H'$ is considered as a probability distribution over $H'$. For any two sets of probability distributions $\mathcal{D}$ and $\mathcal{D}'$ over $H$ and any $\alpha \in [0,1],$ let $\alpha \mathcal{D} + (1-\alpha)\mathcal{D}'$ be the $\alpha$-convex combination of the two sets: $\alpha \mathcal{D} + (1-\alpha)\mathcal{D}'=\{\alpha d + (1-\alpha)d': d \in \mathcal{D}, d' \in \mathcal{D}'\}.$  \medskip

For any function $f: H \rightarrow \mathbb{R}$ and any $d \in \Delta(H),$ define $\mathbb{E}_{d} (f) =  \sum_{h \in H}d(h)f(h)$ as the $d$-evaluation of $f$. For any nonempty $\mathcal{D} \subseteq \Delta(H),$ define $\mathbb{E}_{\mathcal{D}} (f) = \inf_{d \in \mathcal{D}} \mathbb{E}_{d}(f)$ as the $\mathcal{D}$-evaluation of $f$. For any two functions $f$ and $g$, we write $f \ge g$ if $f(h) \ge g(h)$ for all $h \in H.$ For any subset $H' \subseteq H,$  we write $f=^{H'} g$ if $f(h)=g(h)$ for all $h \in H'.$ $f[H']g$ denotes the function that agrees with $f$ on $H'$ and agrees with $g$ on $H\backslash H'.$ For any $\alpha \in [0,1],$ $\alpha f + (1-\alpha)g$ denotes the function satisfying that $(\alpha
f + (1-\alpha)g)(h)=\alpha f(h) + (1-\alpha)g(h)$ for all $h \in H.$ For any partition $\Pi=\{H^1,...,H^n\}$ of $H$, $f$ is said to be measurable with respect to $\Pi$ if $f(h)=f(h')$ whenever $h$ and $h'$ are in the same block of the partition. 
\medskip

Let $H=H_1\times H_2.$ For any $d \in \Delta(H)$ and any $h_2 \in H_2$, if $d(h_1,h_2)>0$ for some $h_1 \in H_1,$ then $d|h_2 \in \Delta(H_1)$ denotes the conditional probability of $d$ on $h_2$: $d|h_2(h'_1)= d(h'_1,h_2)/(\sum_{\hat{h}_1 \in H_1} d(\hat{h}_1, h_2))$ for all $h'_1 \in H_1.$  For any $\mathcal{D} \subseteq \Delta(H)$ and any $h_2 \in H_2$, define the set of conditional distributions of $\mathcal{D}$ on $h_2$ as $\mathcal{D}|h_2=\{d|h_2: d \in \mathcal{D}, \exists h_1 \in H_1 \text{ s.t. } d(h_1,h_2)>0\}.$ For any function $f: H \rightarrow \mathbb{R},$ let $f|h_2$ denote the function mapping from $H_1$ to $\mathbb{R}$ satisfying that $f|h_2(h_1)=f(h_1,h_2), \forall h_1 \in H_1.$
For convenience, we write $h$ for the singleton set $\{h\}$ throughout the paper when there is no confusion.

\subsection{Evaluation Functions and Extended Evaluation Functions}\label{subsec_framework}
In this section, we introduce  evaluation functions  and extended evaluation functions, both of which are assumed to have max-min representations. Each updating rule is defined as a function that maps extended evaluation functions and  signals to evaluation functions. \medskip

Let $S$ be the state space. We assume that $S$ contains at least three states and is finite. Let $\Theta$ be the set of signals. $\Theta$ is countably infinite. Let $\mathbb{K}$ be the payoff space, which is an interval of $\mathbb{R}$ containing a nonempty interior. An act is a function $f: S \rightarrow \mathbb{K}$, and an extended act is a function $f^*: S\times\Theta \rightarrow \mathbb{K}$.   We use $x \in \mathbb{K}$ to denote the constant act as well as the constant extended act that equals $x$ everywhere. Let $\mathcal{F}$ denote the set of all acts and $\mathcal{F}^*$ the set of all extended acts.
\medskip

The DM's priors over $S\times \Theta$ can be revealed from his choices over extended acts. Choices over extended acts can be described by an extended evaluation function. An extended evaluation function is a map $V: \mathcal{F}^* \rightarrow \mathbb{K}$ satisfying that for any $x \in \mathbb{K}$ and any $f^*,g^*, h^* \in \mathcal{F}^*$:\medskip

\textbf{(i) (Identity):} $V(x) = x$. \medskip

\textbf{(ii) (Continuity):} the sets $\{\alpha \in [0,1]: V(\alpha f^*+ (1-\alpha) g^*) \ge V(h^*)\}$ and $\{\alpha \in [0,1]: V(\alpha f^*+ (1-\alpha) g^*) \le V(h^*)\}$ are closed. \medskip

\textbf{(iii) (Certainty Independence):} $\forall \alpha \in (0,1)$, $V(f^*) > V(g^*)$ if and only if $V(\alpha f^*+ (1-\alpha) x) > V(\alpha g^*+ (1-\alpha) x)$. \medskip

\textbf{(iv) (Monotonicity):} $f^*\ge g^*$ implies $V(f^*)\ge V(g^*)$. \medskip

\textbf{(v) (Uncertainty Aversion):}  $V(f^*)=V(g^*)$ implies $V(\frac{1}{2}f^* +\frac{1}{2}g^*) \ge V(f^*)$.\medskip

\textbf{(vi) (Finite Support):} There exists a nonempty and finite subset $\Theta' \subseteq \Theta$ such that $V(f^*)=V(g^*)$ whenever $f^*=^{S\times\Theta'} g^*$. \medskip

\textbf{(vii) (Non-ambiguity over State Space):} If $f^*$ and $g^*$ are measurable with respect to the partition $\{s\times \Theta\}_{s \in S}$ and $V(f^*)=V(g^*)$, then $V(\frac{1}{2}f^* + \frac{1}{2}g^*) = V(f^*)$. \medskip

Let $\mathcal{V}$ be the set of all extended evaluation functions. Conditions (i)-(v) are standard for a max-min expected utility representation.\footnote{See \citet*{jme1989maxmin} for more details.} Together with condition (vi), we know that there exists a unique nonempty, convex and closed set $\mathcal{P} \subseteq \Delta(S\times\Theta')$ such that $V(f^*)=\mathbb{E}_{\mathcal{P}}(f^*)$ for each extended act $f^*.$  $\mathcal{P}$ is said to represent $V.$ Conditions  (vii) ensures that $\mathcal{P}$ induces a unique prior over $S$, i.e., for each $p,p' \in \mathcal{P}$ and each $s \in S,$ $p(s\times\Theta)=p'(s\times\Theta)$. Such a set of priors over $S\times\Theta$ is said to be simple. We discuss more general prior sets in Section \ref{subsec_violation_ISU}.  \medskip

For any $\mathcal{P}$ that is simple, we can decompose it to a prior $\mu$ over $S$ and a set of conditional probabilities $\{c^t(\cdot|s)_{s \in S}\}_{t \in T} \subseteq \left(\Delta(\Theta)\right)^{S}$ satisfying that: 

(1) $\mu(s)=p(s\times\Theta)$ for each $s\in S$ and each $p \in \mathcal{P}$,  

(2) $\forall t \in T$, there exists $p \in \mathcal{P}$ such that $\mu(s)c^t(\theta|s)=p(s,\theta)$, $\forall s \in S$ and  $\forall \theta \in \Theta$, and 

(3) $\forall p \in \mathcal{P}$, there exists $t \in T$ such that $p(s,\theta)= \mu(s)c^t(\theta|s)$, $\forall s \in S$ and  $\forall \theta \in \Theta$.

\noindent We write $\mathcal{P}=(\mu, \{c^t(\cdot|s)_{s \in S}\}_{t \in T})$ if the above conditions hold. Based on the reformulation, a DM who has a simple prior set has no ambiguity over the state space but may have multiple probabilistic interpretations  over the signals, where each $t \in T$ denotes one possible interpretation.
\medskip

After observing some signal $\theta$, the DM's ex-post beliefs can be revealed from his choices over acts.  The DM's choices over acts are described by an evaluation function. An evaluation function is a map $U: \mathcal{F} \rightarrow \mathbb{K}$ satisfying identify, continuity, certainty independence, monotonicity and uncertainty aversion. These conditions ensure that $U$ admits a max-min representation by a unique nonempty, convex and closed set of probability distributions $\mathcal{Q} \subseteq \Delta(S)$. When $\mathcal{Q}=\{\mu^*\}$, we say that $U$ is represented by $\mu^*$. $\mathcal{Q}$ is the DM's set of beliefs over the state space after observing signal $\theta$. Let $\mathcal{U}$ be the set of all evaluation functions. \medskip

We proceed to define updating rules. We only consider signals that happen with non-zero probabilities.\footnote{That is, we do not model how DMs react to unexpected information. For theories of updating events with zero probability, see, for example, \citet*{aer2012unexpected}.} For this purpose, we define non-null signals for an arbitrary extended evaluation function $V$.  $(s,\theta) \in S\times \Theta$ is said to be $V$ null if $V(f^*)=V(g^*)$ for any $f^*,g^* \in \mathcal{F}^*$ satisfying $f^*=^{(S\times \Theta) \backslash (s,\theta)} g^*$. If $(s,\theta)$ is not $V$ null, it is $V$ non-null. It can be easily shown that if $V$ is represented by $\mathcal{P}$, $(s,\theta)$ is $V$ null if and only if for all $p \in \mathcal{P},$ $p(s,\theta) = 0.$ A signal $\theta$ is said to be $V$ non-null if there exists $s \in S$ such that $(s,\theta)$ is $V$ non-null. The set of all $V$ non-null signals is denoted by $\Theta_V.$ $V$ non-null states are defined similarly. In the examples and applications, we omit states and signals that are not non-null.   \medskip

An updating rule is a function $$\Gamma: \bigcup\limits_{V \in \mathcal{V}} \left( \{V\} \times \Theta_V \right) \rightarrow \mathcal{U}.$$ For simplicity, for any $V \in \mathcal{V}$ and $\theta \in \Theta_V$, we write $V_{\theta}$ for  $\Gamma(V,\theta)$. To interpret, the extended evaluation function $V$ characterizes the DM's prior beliefs over the state space and how he interprets signals. Given $V$, an updating rule maps each observed signal to the DM's ex-post evaluation function over acts, which reflects the DM's ex-post beliefs after observing the signal. \medskip

\section{Model}\label{sec_model}
\subsection{CML Updating}\label{subsec_cml_updating}
We introduce CML in this section. Let the DM's prior set over $S\times\Theta$ be a simple set $\mathcal{P}$.  Consider a signal $\theta$ satisfying that $p(s,\theta)>0$ for some $p \in \mathcal{P}$ and $s \in S$.  When signal $\theta$ is observed, CML updating leads to a singleton posterior set $\{\mu_{\theta}\}\subseteq \Delta(S)$, where \begin{equation}\label{eq_d1}
\mu_{\theta}(s) = \frac{\max_{p \in \mathcal{P}}p(s,\theta)}{\sum_{s' \in S} \max_{p \in \mathcal{P}} p(s',\theta) }, \forall s\in S.
\end{equation}
Equivalently, if  $P = (\mu, \{c^t(\cdot|s)_{s \in S}\}_{t \in T})$, $\mu_{\theta}$ can be defined as 
\begin{equation}\label{eq_d2}
\mu_{\theta}(s) = \frac{\mu(s)\max_{t \in \mathcal{T}}c^t(\theta|s) }{\sum_{s' \in S} \mu(s')\max_{t \in \mathcal{T}}c^t(\theta|s')}, \forall s \in S.
\end{equation}
We interpret the above formulation as the DM uses the signal's maximal conditional probability on each state to update his belief.  The DM's CML posterior set is always a singleton and thus is not dilated. With CML,  information does render unambiguous events ambiguous.  \medskip

We note that the CML posterior may not be the Bayes' posterior of any one of the DM's priors over $S\times \Theta$, i.e., $\mu_{\theta}$ may not  be contained in  $\mathcal{P}|\theta$.  This can be seen by the following example.   \medskip

\textbf{Example 1.} Let $S=\{s,s',s''\}$ and $\Theta=\{\theta,\theta'\}.$ The DM's prior set over $S\times \Theta$ is a simple set $\mathcal{P}$, which consists of all convex combinations of $p^1$ and $p^2$. $p^1$ and $p^2$ are shown in Table 1.

\begin{table}[H]
\caption*{\textbf{Table 1}}
\centering
\begin{tabular}{|c|c|c|}
\hline
  $p^1$   & $\theta$  & $\theta'$    \cr\hline
$s$ & $4/15$ & $1/15$     \cr\hline
$s'$ & $4/15$ & $1/15$     \cr\hline 
$s''$ & $1/15$ & $4/15$     \cr\hline 
\end{tabular}
\quad
\begin{tabular}{|c|c|c|}
\hline
 $p^2$     & $\theta$  & $\theta'$    \cr\hline
$s$ & $4/15$ & $1/15$     \cr\hline
$s'$ & $1/15$ & $4/15$     \cr\hline
$s''$ & $4/15$ & $1/15$     \cr\hline 
\end{tabular}
\end{table}

\noindent Since $$\max_{p \in \{p^1,p^2\}} p(s,\theta) = \max_{p \in \{p^1,p^2\}} p(s',\theta) = \max_{p \in \{p^1,p^2\}} p(s'',\theta)=4/15,$$ the DM's CML posterior  is $(1/3, 1/3, 1/3)$ when signal $\theta$ is observed.  Note that $\mathcal{P}|\theta$ consists of all convex combinations of $p^1|\theta$ and $p^2|\theta$, where $p^1|\theta = (4/9,4/9,1/9)$ and $p^2|\theta=(4/9,1/9,4/9)$.  Obviously,  $(1/3,1/3,1/3)$ is not a convex combination of $p^1|\theta$ and $p^2|\theta$. That is, $(1/3, 1/3, 1/3)$ is not contained in $\mathcal{P}|\theta$.  Since the DM updates his belief based on the conditionally most likely scenario for each state,  correlations among different states' conditional probabilities are neglected. Hence, the DM's posterior can be outside of $\mathcal{P}|\theta$. \medskip

The following is the formal definition of CML.

\begin{definition}
An updating rule is CML if for any $V \in \mathcal{V}$ represented by $\mathcal{P}$ and any $\theta \in \Theta_{V}$, the evaluation function $V_{\theta}$ specified by the updating rule can be represented by $\mu_{\theta}$, where  $\mu_{\theta}$ satisfies condition \text{(}\ref{eq_d1}\text{)}.
\end{definition}

\subsection{Axiom ISU}\label{subsec_isu}
In this section, we introduce axiom ISU. This axiom distinguishes CML from existing updating rules. To start with, we introduce two existing updating rules: FB and ML.  Let $V$ be the DM's extended evaluation function represented by $\mathcal{P}$. Let a $V$ non-null signal $\theta$ be observed. \medskip

\textbf{(FB.)} A DM who follows FB updates $\mathcal{P}$ to the posterior set $\mathcal{P}|\theta$. That is, given signal $\theta$, he updates each prior $p$ in $\mathcal{P}$ following Bayes' rule. With FB, the DM's ex-post evaluation function $V_{\theta}$ is represented by the closure of $\mathcal{P}|\theta$.\footnote{If $\min_{p \in \mathcal{P}} p(S\times\theta) >0$, then $\mathcal{P}|\theta$ is nonempty, convex and closed. If $\min_{p \in \mathcal{P}} p(S\times\theta) =0$, then $\mathcal{P}|\theta$ is nonempty and convex but may not be closed.}\medskip

\textbf{(ML.)} A DM who follows ML first selects priors $\mathcal{P}^{\theta,M}$ from $\mathcal{P}$ that maximize the probability of $S\times \theta:$ $$ \mathcal{P}^{\theta,M}=\{p \in \mathcal{P}: p(S\times\theta) \ge p'(S\times\theta), \forall p' \in \mathcal{P}\}.$$ Then, he updates each prior in $\mathcal{P}^{\theta,M}$ following Bayes' rule, which leads to the posterior set $\mathcal{P}^{\theta,M}|\theta$. With ML, the DM's ex-post evaluation function $V_{\theta}$ is represented by $\mathcal{P}^{\theta,M}|\theta$. \medskip

We motivate axiom ISU in an unambiguous environment, in which the DM updates according to Bayes' rule and is an expected utility maximizer. Let $S=\{s,s'\}$ and $\Theta=\{\theta, \theta'\}.$ The DM has a single prior $p$ over $S\times \Theta$ satisfying that $p(s,\theta)=p(s',\theta)=p(s,\theta')=p(s',\theta')=1/4.$  Consider an arbitrary extended act $f^*.$ If the payoff of $f^*$ on $(s,\theta)$ increases by $\epsilon >0,$ its evaluation increases by $\textbf{1/4} \epsilon.$  When $\theta$ is observed, the DM's posterior over $\{s,s'\}$ is $(1/2, 1/2)$. Hence, the DM's ex-post evaluation of the act $f^*|\theta$ increases by $\textbf{1/2} \epsilon$ if the payoff of $f^*$ on $(s,\theta)$ increases by $\epsilon$. \medskip

The key observation from the above example is that the DM becomes more sensitive to payoff changes on $(s,\theta)$ after signal $\theta$ is observed, since the realization of $\theta$ rules out event $S\times (\Theta\backslash \theta)$, which makes payoff changes on $(s,\theta)$ more likely to occur. This argument does not rely on whether the information is ambiguous or not. Based on this observation, we introduce axiom ISU.

\begin{definition}
An updating rule satisfies axiom ISU if for any extended evaluation function $V$, any $\theta \in \Theta_V$ and any $f^*,g^*\in \mathcal{F}^*$ with $g^*(s,\theta) > f^*(s,\theta)$ and $g^*=^{(S\times\Theta) \backslash (s,\theta)} f^*$,  $$V(f^*)=V_{\theta}(f^*|\theta) \text{ implies } V(g^*)\le V_{\theta}(g^*|\theta).$$
\end{definition}

Since $V(f^*)=V_{\theta}(f^*|\theta)$ and the increase of payoff on $(s,\theta)$ affects the DM more after $\theta$ is observed, we have $V(g^*)\le V_{\theta}(g^*|\theta)$. However, the next example shows that axiom ISU is violated by both FB and ML.  \medskip

\textbf{Example 2.} Let $S=\{s, s'\}$ and  $\Theta=\{\theta,\theta'\}$. The DM's prior set over $S\times\Theta$ is  a simple set $\mathcal{P}$, which consists of all convex combinations of $p^1$ and $p^2$. $p^1$ and $p^2$ are shown in Table 2.

\begin{table}[H]\label{table_s_5}
\caption*{\textbf{Table 2}}
\centering
\begin{tabular}{|c|c|c|}
\hline
$p^1$   & $\theta$  & $\theta'$    \cr\hline
$s$ & $1/8$ & $3/8$    \cr\hline
$s'$ & $3/8$ & $1/8$    \cr\hline
\end{tabular}
\quad
\begin{tabular}{|c|c|c|}
\hline
$p^2$   & $\theta$  & $\theta'$    \cr\hline
$s$ & $1/3$ & $1/6$     \cr\hline
$s'$ & $1/8$ & $3/8$     \cr\hline
\end{tabular} 
\quad
\begin{tabular}{|c|c|c|}
\hline
$f^*$   & $\theta$  & $\theta'$    \cr\hline
$s$ & $3$ & $6$     \cr\hline
$s'$ & $2$ & $0$     \cr\hline
\end{tabular}
\quad
\begin{tabular}{|c|c|c|}
\hline
$g^*$   & $\theta$  & $\theta'$    \cr\hline
$s$ & $15/4$ & $6$     \cr\hline
$s'$ & $2$ & $0$     \cr\hline
\end{tabular}
\end{table}

Let $V$ be the DM's extended evaluation function, represented by $\mathcal{P}$. Consider extended acts $f^*$ and $g^*$, as shown in Table 2. We can verify that $$V(f^*)=\mathbb{E}_{\{p^1,p^2\}} (f^*)= \mathbb{E}_{p^2} (f^*) = 9/4,$$
and  
$$V(g^*)=\mathbb{E}_{\{p^1,p^2\}} (g^*) = \mathbb{E}_{p^2} (g^*) = 5/2.$$ Let signal $\theta$ be observed.
\medskip

First, assume that the DM updates according to FB. The DM's posterior set over $S$ consists of all convex combinations of $p^1|\theta= (1/4, 3/4)$ and $p^2|\theta = (8/11, 3/11)$. Hence, a DM who follows FB evaluates $f^*|\theta$ as $\mathbb{E}_{\{p^1|\theta, p^2|\theta\}}(f^*|\theta)$ and evaluates $g^*|\theta$ as $\mathbb{E}_{\{p^1|\theta, p^2|\theta\}}(g^*|\theta).$ We have  $$ \mathbb{E}_{\{p^1|\theta, p^2|\theta\}}(f^*|\theta) = \mathbb{E}_{p^1|\theta}(f^*|\theta) = 9/4,$$
and 
$$\mathbb{E}_{\{p^1|\theta, p^2|\theta\}}(g^*|\theta) = \mathbb{E}_{p^1|\theta}(g^*|\theta) = 39/16.$$

Next, assume that the DM follows ML. The DM selects the prior that maximizes the probability of $S\times\theta,$ which is $p^1.$ Thus, his posterior set is $\{p^1|\theta\}.$ Again, his ex-post evaluations of $f^*|\theta$ and $g^*|\theta$ are given by 
\begin{align*}
& \mathbb{E}_{p^1|\theta}(f^*|\theta) = 9/4 \text{ and } \mathbb{E}_{p^1|\theta}(g^*|\theta) = 39/16.
\end{align*}

Since $\mathbb{E}_{p^2}(f^*)=\mathbb{E}_{p^1|\theta}(f^*|\theta)$ and $\mathbb{E}_{p^2}(g^*)>\mathbb{E}_{p^1|\theta}(g^*|\theta),$  both ML and FB violate axiom ISU. In contrast, CML satisfies axiom ISU, which is implied by the following proposition.

\begin{proposition}\label{prop_isau}
Let $V \in \mathcal{V}$, $\theta \in \Theta_V$ and $V_{\theta}=\Gamma(V,\theta)$, where $\Gamma$ is CML. For any extended acts $f^*$ and $g^*$ and any $(s,\theta) \in S\times \Theta$, if $g^*(s,\theta) > f^*(s,\theta)$  and $g^*=^{(S\times\Theta)\backslash (s,\theta)} f^*$, then $$ V(g^*)-V(f^*)\le V_{\theta}(g^*|\theta) - V_{\theta}(f^*|\theta).$$
\end{proposition}

Proposition \ref{prop_isau} says that CML satisfies a stronger version of axiom ISU. If a DM updates according to CML, he is always more sensitive to payoff changes on $(s,\theta)$ after $\theta$ is observed, no matter whether his ex-ante utility level equals the ex-post level or not.

\section{Characterization}\label{sec_characterization}
In this section, we provide an axiomatic foundation for CML. We provide three axioms to fully characterize CML, two of which are also satisfied by FB. The only axiom that is violated by FB is axiom ISU.  After that, we consider the situation in which a DM's extended evaluation function is not observable. Instead, we observe the DM's prior belief over the state space and posterior belief after observing each signal. We provide a sufficient and necessary condition under which the DM's belief profile can be rationalized by CML, i.e., there exists a set of interpretations of signals such that CML updating leads to the desired posterior  of the DM under each signal.

\subsection{Axiomatic Foundation}\label{subsec_axiom}
First, we define some notations. For any two simple set of priors $\mathcal{P}$ and $\mathcal{P}'$ over $S\times\Theta$ and any   $E \subseteq S\times\Theta,$ 
$\mathcal{P}$ and $\mathcal{P}'$ are said to agree on $E$, denoted by $\mathcal{P} \approx^E \mathcal{P}'$, if 

(1) for any $p \in \mathcal{P}$, there exists $p' \in \mathcal{P}'$ such that for any $E' \subseteq E,$ $p(E')=p'(E'),$ and

(2) for any $p' \in \mathcal{P}'$, there exists $p \in \mathcal{P}$ such that for any $E' \subseteq E,$ $p'(E')=p(E')$.

\noindent By the definition, $\mathcal{P} \approx^E \mathcal{P}'$ means that the two prior sets induce exactly the same set of distributions over $E$. For any two extended evaluation functions $V$ and $V'$ and any  $E \subseteq S\times\Theta$, $V$ and $V'$ are said to agree on $E$, denoted by $V \approx^E V'$, if for any extended act $f^*$ and any payoff $x$, $V(f^*[E]x)=V'(f^*[E]x)$. Obviously, $V$ and $V'$ agree on $E$ if and only if $\mathcal{P}$ and $\mathcal{P}'$ agree on $E$, where $\mathcal{P}$ and $\mathcal{P}'$ represent $V$ and $V'$ respectively.\medskip

For any evaluation function $U$ and any $S' \subseteq S$, $U$ is said to be strictly increasing on $S'$ if for any  $f \in \mathcal{F}$ and any $x,y \in \mathbb{K}$ with $x <y$, $U(x[S']f)<U(y[S']f)$. If $U$ is represented by $\mathcal{Q}$, we can show that $U$ is strictly increasing on $S'$ if and only if $\min_{\mu \in \mathcal{Q}} \mu(S')>0$. We proceed to state the axioms. 
\bigskip

\textbf{Independence of Irrelevant Signals (Axiom IIS).} For any $\{V,W\} \subseteq \mathcal{V}$ and any $\theta \in \Theta_V$, if $V\approx^{S\times\theta}W$, then $V_{\theta} = W_{\theta}.$  \bigskip

Let $V$ be represented by $\mathcal{P}$ and $W$ represented by $\mathcal{P}'$. Note that $V\approx^{S\times\theta}W$ indicates that $\mathcal{P}$ and $\mathcal{P}'$ agree on $S\times\theta.$ Hence, axiom IIS says that the DM's posterior belief after observing signal $\theta$ only depends on the details of the DM's priors on $\theta$ and is irrelevant with the details of other signals.  Obviously, this axiom is satisfied by FB since the FB posterior set consists of each prior's conditional distribution on $S\times \theta$ when $\theta$ is observed. \bigskip

\textbf{Ratio Consistency (Axiom RC).} For any $\{V,W\} \subseteq \mathcal{V}$, any $\theta \in \Theta_V \cap \Theta_W$ and any $\{s,s'\} \subseteq S$ with $s \neq s'$, if $V_{\theta}$ and $W_{\theta}$ are strictly increasing on $\{s,s'\}$, and $V \approx^{\{s,s'\} \times \Theta} W$, then for any $\{y,z,w\} \subseteq \mathbb{K},$ $V_{\theta}(y[s]z[s']w) = w$ if and only if $W_{\theta}(y[s]z[s']w)= w.$\bigskip

Again, let $V$ be represented by $\mathcal{P}$ and $W$  represented by $\mathcal{P}'.$
The primitive conditions of axiom RC indicate that $\mathcal{P}$ and $\mathcal{P}'$ agree on $\{s,s'\}\times \Theta$.
Axiom RC says that how the DM trades off payoffs on $s$ and $s'$ after observing signal $\theta$ is completely determined by the details of the DM's priors on the two states. More concretely, let $V_{\theta}$ be represented by $\mathcal{Q}_{\theta}$ and $W_{\theta}$ represented by $\mathcal{Q}'_{\theta}$. Since $V_{\theta}$ and $W_{\theta}$ are strictly increasing on $\{s,s'\}$, each posterior in $\mathcal{Q}_{\theta} \cup \mathcal{Q}'_{\theta}$ assigns a positive probability to $\{s,s'\}$. The condition that $V_{\theta}(y[s]z[s']w) = w$ if and only if $W_{\theta}(y[s]z[s']w)= w$ is equivalent to that 
$$ \max_{\mu \in \mathcal{Q}_{\theta}} \frac{\mu(s)}{\mu(s')} = \max_{\mu' \in \mathcal{Q}'_{\theta}} \frac{\mu'(s)}{\mu'(s')},$$
and  
$$\min_{\mu \in \mathcal{Q}_{\theta}} \frac{\mu(s)}{\mu(s')} = \min_{\mu' \in \mathcal{Q}'_{\theta}} \frac{\mu'(s)}{\mu'(s')},$$ 

\noindent where the maximization and minimization are allowed to be positive infinite as the denominator can be equal to zero.  We note that FB also satisfies this axiom: with FB, the ratios of the two states' posterior probabilities are completely determined by their ex-ante probability ratios on the realized signal.\footnote{We prove that FB satisfies axiom RC  in Appendix B.} The last axiom is axiom ISU.\bigskip

\textbf{increased Sensitivity after Updating (Axiom ISU).} For any $V \in \mathcal{V}$, any $\theta \in \Theta_V$ and any $\{f^*,g^*\} \subseteq \mathcal{F}^*$ satisfying $f^*(s,\theta) < g^*(s,\theta)$ and $f^* =^{(S\times\Theta) \backslash (s,\theta)} g^*,$  $V(f^*)=V_{\theta}(f^*|\theta)$ implies $V(g^*)\le V_{\theta}(g^*|\theta)$. \bigskip

The three axioms fully characterize CML.  \medskip

\begin{theorem}\label{thm_main}
An updating rule is CML if and only if it satisfies axioms IIS, RC and ISU.
\end{theorem}

We sketch the proof of the sufficiency part of the theorem. Consider an extended evaluation function $V$ and let it be represented by $\mathcal{P}=(\mu, \{c^t(\cdot|s)_{s \in S}\}_{t \in T})$. Recall that $\mu$ is the DM's prior over the state space, and $\{c^t(\cdot|s)_{s \in S}\}_{t \in T}$ are the DM's interpretations of signals. Consider some $\theta \in \Theta_V$ and let $V_{\theta}$ be represented by $\mathcal{Q}_{\theta} \subseteq \Delta(S)$. \medskip

The most important step is to show that $\min_{\mu^* \in \mathcal{Q}_{\theta}} \mu^*(s) \ge \mu(s)$ if $\mu(s)>0$ and $\max_{t\in T}c^t(\theta|s)=1$. To see this, consider an extended act $f^*$ satisfying that 

(1) $f^*(s,\theta)=0,$ 

(2) $f^*(s,\theta')=1$ for each $\theta' \neq \theta$ and 

(3) $f^*(\hat{s},\hat{\theta})=0$ for each $(\hat{s},\hat{\theta}) \in (S\backslash s)\times\Theta$.\footnote{Here, we assume that both $0$ and $1$ are in the payoff space $\mathbb{K}$. $0$ and $1$ can be replaced by any payoffs $x$ and $y$ satisfying $x < y$.}

\noindent The $\mathcal{P}$-evaluation of $f^*$, i.e., $V(f^*)$, is $0$ since $\max_{t\in T}c^t(\theta|s)=1$, which indicates that the minimal probability of $s \times (\Theta \backslash \theta)$ is zero. For $\epsilon \in (0,1)$, an $\epsilon$ increase of payoff on $(s,\theta)$ increases the $\mathcal{P}$-evaluation of $f^*$ by $\mu(s) \epsilon$. When $\theta$ is observed, the act $f^*|\theta$ yields payoff $0$ constantly. Hence, the $\mathcal{P}$-evaluation of $f^*$ equals to the $\mathcal{Q}_\theta$-evaluation of $f^*|\theta$. Note that an $\epsilon$ increase of payoff on $(s,\theta)$ for $f^*$ increases the $\mathcal{Q}_\theta$-evaluation of $f^*|\theta$ by $\min_{\mu^* \in \mathcal{Q}_{\theta}} \mu^*(s)\epsilon$. Axiom ISU implies that $\min_{\mu^* \in \mathcal{Q}_{\theta}} \mu^*(s) \ge \mu(s).$ \medskip

Given the above observation, the rest of the proof can be illustrated through the following simple example, in which there is no ambiguity. General cases can be shown similarly. \medskip

Let $S=\{s, s', s''\}$ and $\Theta=\{\theta, \theta'\} \cup \{\theta_i\}_{i=1}^{+\infty}$. Consider $\mathcal{P}=(\mu, \{c(\cdot|\hat{s})_{\hat{s}\in S}\})$, where $\mu=(1/3,1/3,1/3)$ and $c(\theta|s)=c(\theta'|s)=c(\theta|s')=c(\theta'|s')=c(\theta|s'')=c(\theta'|s'')=1/2.$ 
When $\theta$ is observed, we want to show that for each one of the DM's  posteriors, the probability ratio between states $s$ and $s'$ is equal to    
\begin{equation}\label{eq_desired_ratio}
\frac{\mu(s)c(\theta|s)}{\mu(s')c(\theta|s')}.
\end{equation}

Consider the extended evaluation function $V'$, represented by $\mathcal{P}'=(\mu',\{c'(\cdot|\hat{s})_{\hat{s} \in S}\})$ where $$\mu'(s) =  \mu(s)c(\theta|s) =1/6, \mu'(s')=\mu(s')(\theta|s') =1/6,  \mu'(s'')=2/3,$$ $$ c'(\theta|s)=c'(\theta|s')=1, \text{ and }c'(\theta|s'')= 1/4.$$
We can verify that $\mathcal{P}\approx^{S\times\theta} \mathcal{P}'$. By axiom IIS, $V_{\theta}=V'_{\theta}$. Let $V'_{\theta}$ be represented by $\mathcal{Q}'_{\theta}$. We have $\mathcal{Q}'_{\theta}=\mathcal{Q}_\theta$. \medskip

Consider another extended evaluation function $V''$, represented by $\mathcal{P}''=( \mu'', \{c''(\cdot|\hat{s})_{\hat{s} \in S}\})$ where  $$\mu''(\hat{s}) = \mu'(\hat{s}), \forall \hat{s} \in S,$$
and 
$$c''(\theta|s)=c''(\theta|s')= c''(\theta|s'')= 1.$$
Obviously, $\mathcal{P}' \approx^{\{s,s'\}\times \Theta} \mathcal{P}''$. Note that for each $\hat{s} \in S$, $c''(\theta|\hat{s})=1$. Hence, if $V''_{\theta}$ is represented by $\mathcal{Q}''_{\theta}$, we have $\min_{\mu^* \in \mathcal{Q}''_{\theta}}\mu^*(\hat{s}) \ge \mu''(\hat{s})$ for all $\hat{s} \in S$. This indicates that $\mathcal{Q}''_{\theta}=\{\mu''\}$. Since $\mathcal{P}' \approx^{\{s,s'\}\times \Theta} \mathcal{P}''$, by axiom RC, we know that for each $\mu^* \in \mathcal{Q}'_{\theta}$($\mathcal{Q}_\theta$), the ratio between $\mu^*(s)$ and $\mu^*(s')$ is  equal to the ratio between $\mu''(s)$ and $\mu''(s')$, which is the desired ratio (\ref{eq_desired_ratio}).  \medskip

For completeness, we show that any two out of the three axioms are not sufficient for an updating rule to be CML in Appendix B.

\subsection{CML Rationalizability}\label{subsec_rational}
In this section, we  study the situation in which the DM's extended evaluation function cannot be observed. Let the DM's ex-ante evaluation function be $U$. $U$ describes the DM's choices over acts before any information. Let $U$ be represented by $\mu \in \Delta(S).$ We assume throughout this section that $\mu \in \Delta^o(S).$\footnote{If $\mu(s)=0$, we can simply delete state $s$ and define the state space to be $S\backslash\{s\}$. If any posterior of the DM assigns positive probability to $s$, we can directly reject the hypothesis that the DM follows CML.} The DM's choices over acts under signal $\theta$ is characterized by the ex-post evaluation function $U_{\theta}.$ Let $\{U_{\theta}\}_{\theta \in \Theta^*}$ be the DM's set of ex-post evaluation functions, where $\Theta^*$ is a finite set of signals. $U_{\theta}$ is represented by $\mu_{\theta} \in \Delta(S).$\footnote{Again, if $U_{\theta}$ is represented by a non-singleton set $\mathcal{Q}_{\theta}$, we can  reject the hypothesis that the DM follows CML.}  Given the profile $(\mu, \{\mu_{\theta}\}_{\theta \in \Theta^*}),$ we want to provide a sufficient and necessary condition under which the ex-post beliefs of the DM are updated according to CML under some set of interpretations of signals.

\begin{definition}
$(\mu, \{\mu_{\theta}\}_{\theta \in \Theta^*})$ is CML rationalizable if there exists a finite set of interpretations $\{c^t(\cdot|s)_{s\in S}\}_{t \in T}$ over $\Theta^*$ such that for each $\theta \in \Theta^*,$  
\begin{equation}\label{eq_nonnull}
\sum_{s' \in S} \mu(s') \max_{t \in T} c^t(\theta|s') > 0,
\end{equation}
and 
\begin{equation}\label{eq_cml_rational}
\mu_{\theta} (s) = \frac{ \mu(s) \max_{t \in T} c^t(\theta|s) }{\sum_{s' \in S} \mu(s') \max_{t \in T} c^t(\theta|s')}, \forall s \in S.
\end{equation}
\end{definition}\medskip

Condition (\ref{eq_nonnull}) requires that all observed signals are not unexpected. Condition (\ref{eq_cml_rational})  says that the DM updates according to CML. The following example shows that not all belief profiles are CML rationalizable. \medskip

\textbf{Example 3.} Let $S=\{s,s',s''\}$ and $\Theta^*=\{\theta,\theta'\}$. Suppose that the DM's prior over $S$ is $\mu=(1/3, 1/3, 1/3)$. His posterior is $\mu_{\theta}= (1/6, 1/2, 1/3)$ under signal $\theta$ and $\mu_{\theta'}= (1/6, 1/2, 1/3)$ under signal $\theta'$. We argue that $(\mu, \{\mu_{\theta}, \mu_{\theta'}\})$ is not CML rationalizable. The reason is that the posterior probabilities of state $s$ are too small compared to the posterior probabilities of state $s'.$ When signal $\theta$ is realized, the conditional likelihoods ratio used for updating between $s$ and $s'$  is $1: 3$. Since the conditional maximum likelihood of $\theta$ on $s'$ is at most $1$, the conditional maximum likelihood of $\theta$ on $s$ is at most $1/3.$ By a similar argument, the conditional maximum likelihood of $\theta'$ on $s$ is at most $1/3.$ Obviously, this is impossible, since it indicates that the summation of the conditional probabilities of $\theta$ and $\theta'$ on $s$ is strictly less than one. \medskip

The next theorem completely characterizes the set of CML rationalizable belief profiles. The characterization exactly rules out the situation in  Example 3: conditional on any state $s$, the summation of the probabilities of all signals cannot be strictly less than one.

\begin{theorem}\label{thm_rational}
$(\mu, \{\mu_{\theta}\}_{\theta \in \Theta^*})$ is CML rationalizable if and only if for each $s \in S,$ $$\sum_{\theta \in \Theta} \frac{\mu_{\theta}(s)}{\mu(s)}   \left( \max_{s' \in S'}\frac{\mu_{\theta}(s')}{\mu(s')} \right)^{-1} \ge 1.$$
\end{theorem}

\section{Application}\label{sec_application}
In this section, we give three applications of CML. First, we show that CML is divisible, i.e., CML updating is unaffected by the order independent signals arrive.  Second, we apply CML to study how DMs update with signals of unknown accuracy. We show that CML predicts under-reaction to such signals and illustrate how DMs learn with such signals. Finally, we apply CML to study an information design problem where a designer can introduce ambiguous information structures to affect the action taken by an agent. We show that if the agent follows CML, the designer can almost achieve the maximal payoff.

\subsection{Divisibility}\label{subsec_divisibility}
An important question in the literature of belief updating is whether DMs' beliefs are affected by the order independent signals arrive. An updating rule is divisible, or path-independent, if the posterior set of the DM is unaffected by the order of independent signals. \citet*{wp2019divisible} gives the representations of divisible rules when there is no ambiguity. In this section, we show that CML satisfies divisibility. \medskip

For simplicity, let $S$, $\Theta$ and $\Theta'$ be finite. $\Theta$ and $\Theta'$ are two different sets of signals. Let the DM's prior over $S$ be $\mu$. Let the DM's set of interpretations over $\Theta$ be $\{ c^t(\cdot|s)_{s \in S} \}_{t\in T}$ and set of interpretations over $\Theta'$ be $\{ c^{t'}(\cdot|s)_{s \in S} \}_{t' \in T'}.$ The two signal sets $\Theta$ and $\Theta'$ are independent: the DM's set of interpretations over $\Theta \times \Theta'$ is given by $\{c^{(t,t')}(\cdot|s)_{s \in S}\}_{(t,t') \in T\times T'}$ where for each $(t,t') \in T\times T'$, $c^{(t,t')}( \theta,\theta' |s)=c^t(\theta|s)\cdot c^{t'}(\theta'|s)$ for each $\theta \in \Theta$ and $\theta' \in \Theta'.$ \medskip

Fix the two independent signal sets. We argue that CML updating is path-independent. First, assume that signals $\theta$ and $\theta'$ arrive simultaneously. The DM's posterior $\mu_{\theta,\theta'}$ is given by
$$\mu_{\theta,\theta'}(s)  =\frac{\mu(s) \max_{(t,t') \in T\times T'} c^{(t,t')}(\theta, \theta'|s)}{\sum_{s' \in S}  \mu(s') \max_{(t,t') \in T\times T'} c^{(t,t')}(\theta, \theta'|s')}, \forall s \in S.$$
\noindent Since $c^{(t,t')}(\theta,\theta'|s)=c^t(\theta|s)\cdot c^{t'}(\theta'|s)$, we have
$$\mu_{\theta,\theta'}(s) = \frac{\mu(s) \max_{t \in T} c^{t}(\theta|s) \max_{t' \in T'} c^{t'}(\theta'|s)}{\sum_{s' \in S}  \mu(s') \max_{t \in T} c^{t}(\theta|s') \max_{t' \in T'} c^{t'}(\theta'|s')}, \forall s \in S.$$
Based on the updating formula, the conditional likelihoods used for updating is the multiplication of the conditional maximum likelihoods of the two signals on the each state. \medskip

Next, consider the case where signal $\theta$ is observed first. The DM updates his prior $\mu$ to $\mu_{\theta}:$
\begin{align*}
\mu_{\theta}(s) =  \frac{\mu(s) \max_{t \in T} c^{t}(\theta|s) }{\sum_{s' \in S}  \mu(s') \max_{t \in T} c^{t}(\theta|s')  }, \forall s\in S.
\end{align*}
Then, he observes signal $\theta'$ and updates $\mu_{\theta}$ to $(\mu_{\theta})_{\theta'}:$
\begin{align*}
(\mu_{\theta})_{\theta'}(s) =  \frac{\mu_{\theta}(s) \max_{t' \in T'} c^{t'}(\theta'|s) }{\sum_{s' \in S}  \mu_{\theta}(s') \max_{t' \in T'} c^{t'}(\theta'|s') }, \forall s \in S.
\end{align*}
Obviously, $\mu_{\theta,\theta'}$ and $(\mu_{\theta})_{\theta'}$ are the same. Similarly, if signal $\theta'$ arrives before $\theta$, the DM's posterior remains to be $\mu_{\theta,\theta'}$.  Hence, CML is divisible.

\subsection{Information with Ambiguous Accuracy}\label{subsec_accuracy}
SO19 and L19 study how DMs react to information of ambiguous accuracy. Some of their experimental findings are inconsistent with FB and ML under the max-min expected utility framework. For example, the experimental evidence of SO19 suggests that ambiguity averse DMs may not dilate their prior when processing ambiguous information. In addition to the non-dilation property, we show that CML can accommodate another  experimental finding by L19: under-reaction to ambiguous information. We then discuss how DMs learn in the long run with CML when information has ambiguous accuracy.\medskip

We introduce the framework. Let $S=\{s_1, s_2\}$ be the state space and $\Theta = \{\theta_1,\theta_2\}$ the signal space. There are two possible levels of accuracy of the signals: $H$ and $L$. We require that $H,L \in (0,1)$, $H > L$ and $H + L \ge 1.$ The DM's set of interpretations of signals are given by $\{c^H, c^L\}$ where $c^H(\theta_1|s_1)=c^H(\theta_2|s_2)=H$ and $c^L(\theta_1|s_1)=c^L(\theta_2|s_2)=L.$    $H+L \ge 1$ ensures that  the likelihood of $s_i$ is weakly increased after $\theta_i$ is observed for each $i \in \{1,2\}$. Otherwise, we can swap the labels of the two signals. \medskip

The first observation is that when $H=1-L$, the DM will maintain his prior over $S$ upon receiving any signal with CML. Since given any signal $\theta_i$, the conditional maximum likelihood of $\theta_i$ on $s_i$ is $H$ and the conditional maximum likelihood of $\theta_i$ on $s_{j}$ ($j \neq i$) is $1-L$. The two conditional likelihoods cancel out.  This coincides with the experimental finding of SO19 that ambiguous averse DMs do not adjust their value of bets after such signals.  \medskip

Next, we introduce the definition of under-reaction to unambiguous information by L19. Consider acts $f$
and $g$ satisfying that $f(s_1)=g(s_2)=1$ and $f(s_2)=g(s_1)=0.$ Consider another conditional distribution $c(\cdot|s)_{s \in S}$ where $$c(\theta_1|s_1)=c(\theta_2|s_2)=\frac{H+L}{2}.$$ 
A DM \textit{under-reacts} to ambiguous information if when signal $\theta_1$ is observed, his ex-post evaluation of $f$ under the interpretations $\{c^H, c^L\}$ is strictly lower than that under $\{c\},$ and his ex-post evaluation of $g$ under the interpretations $\{c^H, c^L\}$ is strictly higher than that under $\{c\}.$ To interpret, observing  signal $\theta_1$ is good news for $f$ and bad news for $g$. Hence, after observing $\theta_1$, the DM increases his evaluation of $f$ and decreases his evaluation of $g$. If he under-reacts to ambiguous information, he increases less for his evaluation of $f$ and decreases less for his evaluation of $g$ when the interpretation of signals has multiple possibilities. \medskip

CML predicts under-reaction to ambiguous information when $H+L>1.$ To see this, let the DM's prior over $\{s_1,s_2\}$ be $\mu=(\alpha, 1-\alpha),$ where $\alpha \in (0,1).$ Assume that signal $\theta_1$ is observed. Given the set of interpretations $\{c^H, c^L\}$, the DM's CML posterior $\mu_{\theta_1}$ satisfies
\begin{align*}
\mu_{\theta_1}(s_1) &= \frac{\alpha H}{\alpha H +(1-\alpha)(1-L)},  \\
\mu_{\theta_1}(s_2) &= \frac{(1-\alpha)(1-L)}{\alpha H +(1-\alpha)(1-L)}.
\end{align*}
Given the unambiguous interpretation $\{c\}$, the DM's CML posterior $\mu^*_{\theta_1}$ satisfies
\begin{align*}
 \mu^*_{\theta_1}(s_1) &= \frac{\alpha \frac{H+L}{2}}{\alpha \frac{H+L}{2} +(1-\alpha)(1-\frac{H+L}{2})},\\
\mu^*_{\theta_1}(s_2) &= \frac{(1-\alpha)(1-\frac{H+L}{2})}{\alpha \frac{H+L}{2} +(1-\alpha)(1-\frac{H+L}{2})}.
\end{align*}
To see that the DM always under-reacts to ambiguous information, note that the two conditional likelihoods ratios satisfy
\begin{align*}
\frac{H}{1-L} -  \frac{(H+L)/2}{1-(H+L)/2} =   & \frac{H-L-H^2+L^2}{(1-L)(2-H-L)}\\
 =  & \frac{(H-L)(1-H-L)}{(1-L)(2-H-L)} \\
 < & 0.
\end{align*} 
Hence, the DM updates more when the information is unambiguous. The same argument applies for signal $\theta_2$. \medskip

We end this section by discussing the implications of CML on learning. Consider the situation in which the DM's set of interpretations of signals consists of $c^H$ and $c^L.$ Assume that  $H+L >1$. Suppose that the true signal generating process is given by some $c^*$ where $c^*(\theta_1|s_1)=c^*(\theta_2|s_2)=\lambda \in (0,1).$ Fix the true state of the world. At each period, a signal from $\{\theta_1, \theta_2\}$ is  generated according to $c^*$ independently. CML  has simple predictions on learning in such an environment. 
\medskip

First, note that the DM always updates according to the conditional likelihood ratio $\frac{H}{1-L}$. Hence, the DM updates his belief as if the accuracy of signals is $\frac{H}{H+1-L} > 1/2$. If $c^*$ is weakly consistent with the DM's perception of the information, i.e., $\lambda > 1/2,$ the DM will finally learn the true state of the world. In contrast, if  $\lambda < 1/2,$ the DM will finally learn the wrong state of the world. Hence, even with ambiguous information, as long as the DM's interpretations of signals are biased towards the true signal generating process, he will finally learn the true state if he updates according to CML.

\subsection{Information Design}\label{subsec_information_design}
The problem of information design is first studied by \citet*{aer2011persuasion}, in which information structures are unambiguous. \citet*{jet2019pers} study the case where the information designer can use ambiguous information structures, and the agent follows FB. In this section, we consider ambiguous information structures and assume that the agent follows CML. \medskip

Let a finite set $S$ be the state space. A designer and an agent have a common prior $\mu \in \Delta^o(S).$ Let a finite set $A$ be the set of all actions. The designer can design information structures for the agent. After observing a signal, the agent chooses an action. The utilities of the designer and the agent depend on both the true state and the action taken by the agent. Let $v: S\times A \rightarrow \mathbb{R}$ be the designer's utility function. Let $u: S \times A \rightarrow \mathbb{R}$ be the agent's utility function. \medskip

Since the agent follows CML, the posterior set of the agent is always a singleton.
For any $\mu^* \in \Delta(S),$ let $\mathcal{BR}(\mu^*)$ be the set of actions that maximize the expected utility of the agent under $\mu^*,$ i.e., $$\mathcal{BR}(\mu^*)=\{a \in A:\sum_{s\in S} \mu^*(s)u(s,a) \ge \sum_{s\in S} \mu^*(s)u(s,a'), \forall a' \in A\}.$$
Let $A^* \subseteq A$ be the set of all optimal actions of the agent, i.e., $$A^* = \bigcup_{\mu^* \in \Delta(S)} \mathcal{BR}(\mu^*).$$ We impose the following assumption on $A^*.$  \begin{assumption}\label{assume_generic}
For each $a \in A^*,$ there exists $\mu^* \in \Delta^o(S)$ such that $a \in \mathcal{BR}(\mu^*).$
\end{assumption}
\noindent The assumption says that the agent's optimal actions can be achieved by posteriors in the interior of $\Delta(S).$ Note that this assumption is generically true. \medskip

We next define ambiguous information structures. An ambiguous information structure is a tuple $(\Theta, \{c^{t}(\cdot|s)_{s \in S}\}_{t\in T})$ where $\Theta$ is a finite set of signals and $\{c^{t}(\cdot|s)_{s \in S}\}_{t\in T} \subseteq (\Delta(\Theta))^{S}$ is a finite set of signal generating systems. We require that for each $\theta \in \Theta,$ $\max_{t \in T}c^t(\theta|s)>0$ for some $s \in S$. Let $\mu_{\theta}$ be the CML posterior of $\mu$ under this information structure when signal $\theta$ is observed, i.e., $\mu_\theta$ satisfies condition (\ref{eq_d2}).

Consider an ambiguous information structure $(\Theta, \{c^{t}(\cdot|s)_{s \in S}\}_{t\in T} )$ and a set of actions $\{a^{\theta}\}_{\theta \in \Theta},$ we say that $G=(\Theta, \{c^{t}(\cdot|s)_{s \in S}\}_{t\in T}, \{a^{\theta}\}_{\theta \in \Theta})$ is implementable if

1. For each $\theta \in \Theta,$ $a^{\theta} \in \mathcal{BR}(\mu_{\theta}).$

2. For any $t,t' \in T,$ $$\sum_{s \in S}\sum_{\theta \in \Theta} \mu(s)c^t(\theta|s)v(s,a^{\theta}) = \sum_{s \in S}\sum_{\theta \in \Theta} \mu(s)c^{t'}(\theta|s)v(s,a^{\theta}).$$

\noindent Condition 1 says that $a^\theta$ is the agent's optimal action when observing signal $\theta$.  Condition 2 ensures the commitment power of the designer: she has no incentive to pick any particular signal generating system since under each signal generation system, her expected payoff is the same. If $G=(\Theta, \{c^{t}(\cdot|s)_{s \in S}\}_{t\in T}, \{a_{\theta}\}_{\theta \in \Theta})$ is implementable, let $v^*(G)$ denote the designer's expected payoff under $G,$ i.e., $$v^*(G)=\sum_{s \in S}\sum_{\theta \in \Theta} \mu(s)c^t(\theta|s)v(s,a^{\theta}), \forall t\in T.$$
Ideally, the designer's maximal payoff is given by $$\sum_{s \in S} \left( \mu(s)\max_{a \in A^*} v(s,a) \right).$$
That is, under each state of the world, the agent takes the designer's optimal action from the feasible action set $A^*$. The next theorem states that the maximal payoff of the designer can be almost achieved with suitable information structures.

\begin{theorem}\label{thm_indesign}
Suppose that assumption \ref{assume_generic} holds. For any $\epsilon >0$, there exists an ambiguous information structure $(\Theta, \{c^{t}(\cdot|s)_{s \in S}\}_{t\in T})$ and a set of actions $\{a^{\theta}\}_{\theta \in \Theta}$ such that $G=(\Theta, \{c^{t}(\cdot|s)_{s \in S}\}_{t\in T}, \{a^{\theta}\}_{\theta \in \Theta})$ is implementable and $$v^*(G)> \sum_{s \in S} \left( \mu(s)\max_{a \in A^*} v(s,a) \right) -\epsilon. $$
\end{theorem}

We give a simple example to illustrate the idea of Theorem \ref{thm_indesign}. Let $S=\{s_1, s_2\}$ and $A^*=\{a_1, a_2\}$. $a_1$ is the designer's optimal action on state $s_1$.  $a_2$ is the designer's optimal action on $s_2$. The common prior of the designer and the agent is $\mu=(1/2, 1/2)$.  The agent's optimal action is $a_1$ if his posterior is $\mu_1=(1/3, 2/3)$; the agent's optimal action is $a_2$ if his posterior is $\mu_2=(2/3, 1/3)$. \medskip

Consider the following ambiguous information structure. There are $20000$ signals $\Theta=\{\theta^{j,l}\}_{j\in\{1,...,1000\}, l \in \{1,2\}}$ and $20000$ signal generating systems $\{c^{j,l}\}_{j\in\{1,...,1000\}, l \in \{1,2\}}$. Our target is that for $l \in \{1,2\}$, when signal $\theta^{j,l}$ is realized, the agent's CML posterior is $\mu_l$. Thus, the agent chooses action $a_l$ at signal $\theta^{j,l}$ for any  $j$. Moreover, we want the agent's posterior at signal $\theta^{j,l}$ to be induced by $c^{j,l}$ for each $j$ and each $l$. We illustrate the construction of 
$c^{j,1}$ for some $j$ as an example. Let $$c^{j,1} (\theta^{j,1}|s_1) =0.01, \text{ and } c^{j,1} (\theta^{j,1}|s_2) =0.02.$$ 
The two conditional probabilities indeed induce the posterior $\mu_1=(1/3, 2/3)$ given the prior $\mu=(1/2,1/2)$. We want to allocate the rest of the conditional probabilities correctly, i.e., we want the rest of the conditional probabilities on each state to be allocated to the signals that induce the designer's optimal action on the state. Let 
$$c^{j,1} (\theta^{j',1}|s_1) =\frac{1-0.01}{10000-1}, \forall j' \neq j,$$
$$c^{j,1} (\theta^{j^*,2}|s_2) =\frac{1-0.02}{10000}, \forall j^*,$$
$$c^{j,1} (\theta^{j^*,2}|s_1) =0, \forall j^*, \text{ and } c^{j,1} (\theta^{j',1}|s_2)=0, \forall j' \neq j.$$
With $c^{j,1}$, conditional on $s_1$, the agent chooses action $a_1$ with  probability $1$. Condition on $s_2$, the agent chooses action $a_2$ with probability $0.98$. Other signal generating systems can be constructed similarly. We need many signals to ensure the conditional probability of $\theta^{j,l}$ to be very small under $c^{j',l'}$ if $(j,l) \neq (j',l')$ so that only $c^{j,l}$ plays a role for controlling the agent's posterior at signal $\theta^{j,l}$. Note that the only payoff loss under $c^{j,1}$ is when $(s_2, \theta^{j,1})$ is realized. However, we can let the payoff loss be smaller by adding more signals and shrinking the conditional probability $c^{j,1} (\theta^{j,1}|s_2)$.  \medskip

We note that  the optimal payoff is usually not achievable. For example, assume that $a \in A^*$ is the the designer's unique optimal action at state $s$ and is not an optimal action of the designer at any other state. Assume further that the agent chooses $a$ only if the probability of state $s'$ ($s' \neq s$) is greater than half. In this case, the designer must incur a payoff loss in order to induce action $a$, where the payoff loss comes from the situation in which $a$ is chosen at state $s'$.

\section{Discussion}\label{sec_discussion}
\subsection{A More General Framework: Violation of Axiom ISU}\label{subsec_violation_ISU}
A natural extension of our framework is to allow the DM to have multiple priors over the state space. However, we show that axiom ISU will be violated by any updating rule. The only assumption we impose on the updating rule is that if $(s,\theta)$ has zero prior probability, then $s$ has zero posterior probability after signal $\theta$ is observed. Consider the following example.\medskip

\textbf{Example 4.} Let $S=\{s_1, s_2, s_3, s_4\}$ and $\theta=\{\theta_1, \theta_2\}.$ The DM's prior set over $S\times\Theta$ is $\mathcal{P}$, consisting of all convex combinations of $p^1$ and $p^2$, as shown in Table 3. Note that $\mathcal{P}$ is not simple since it does not induce a unique prior over $S$.

\begin{table}[H]
\caption*{\textbf{Table 3}}
\centering
\begin{tabular}{|c|c|c|}
\hline
  $p^1$   & $\theta_1$  & $\theta_2$    \cr\hline
$s_1$ & $0.1$ & $0$     \cr\hline
$s_2$ & $0.7$ & $0$     \cr\hline 
$s_3$ & $0$ & $0.1$     \cr\hline 
$s_4$ & $0$ & $0.1$     \cr\hline 
\end{tabular}
\quad
\begin{tabular}{|c|c|c|}
\hline
$p^2$   & $\theta_1$  & $\theta_2$    \cr\hline
$s_1$ & $0.6$ & $0$     \cr\hline
$s_2$ & $0.1$ & $0$     \cr\hline 
$s_3$ & $0$ & $0.2$     \cr\hline 
$s_4$ & $0$ & $0.1$     \cr\hline 
\end{tabular}
\end{table}

Let $\theta_1$ be observed. Since only $(s_1,\theta_1), (s_2,\theta_1), (s_3, \theta_2)$ and $(s_4, \theta_2)$ have non-zero prior probabilities, the DM's posterior set is given by some $\mathcal{Q} \subseteq \Delta(\{s_1,s_2\})$. Assume without loss of generality that the payoff space is $\mathbb{R}.$ Consider two extended acts $f^*$ and $g^*:$ $$f^*(s_1,\theta_1)=0, f^*(s_2,\theta_1)=1, f^*(s_3,\theta_2)=10, f^*(s_4,\theta_2)=x,$$
$$g^*(s_1,\theta_1)=1, g^*(s_2,\theta_1)=0, g^*(s_3,\theta_2)=-10, g^*(s_4,\theta_2)=y,$$
where $x$ and $y$ will be determined later on.  The $\mathcal{P}$-evaluations of $f^*$ and $g^*$ are given by $$\mathbb{E}_{\{p^1,p^2\}}(f^*) = \mathbb{E}_{p^1}(f^*)=  1.7+0.1 x,$$
$$\mathbb{E}_{\{p^1,p^2\}}(g^*) = \mathbb{E}_{p^2}(g^*) = -1.4+ 0.1 y.$$
After $\theta_1$ is observed, the ex-post evaluations are given by
$$\mathbb{E}_{\mathcal{Q}}(f^*|\theta)=\min_{\mu^* \in \mathcal{Q}} \mu^*(s_2),$$
$$\mathbb{E}_{\mathcal{Q}}(g^*|\theta)=\min_{\mu^* \in \mathcal{Q}} \mu^*(s_1).$$
Let $\epsilon >0$ be small enough. Increasing the payoff of $f^*$ on $(s_2,\theta_1)$ by $\epsilon$ increases its $\mathcal{P}$-evaluation by $p^1(s_2,\theta_1) \epsilon$ and increases the $\mathcal{Q}$-evaluation of $f^*|\theta_1$  by $\min_{\mu^* \in \mathcal{Q}} \mu^*(s_2) \epsilon$. Similarly, increasing the payoff of $g^*$ on $(s_1,\theta_1)$ by $\epsilon$ increases its $\mathcal{P}$-evaluation by $p^2(s_1,\theta_1) \epsilon$ and increases the $\mathcal{Q}$-evaluation of $g^*|\theta_1$  by $\min_{\mu^* \in \mathcal{Q}} \mu^*(s_1) \epsilon$. We can pick $x$ and $y$ such that $$ \mathbb{E}_{p^1}(f^*) = \mathbb{E}_{\mathcal{Q}}(f^*|\theta),$$
$$ \mathbb{E}_{p^2}(g^*) = \mathbb{E}_{\mathcal{Q}}(g^*|\theta).$$
By axiom ISU, we conclude that 
$$ \min_{\mu^* \in Q}\mu^*(s_2) \ge  p^1(s_2,\theta_1)=0.7,$$
and
$$ \min_{\mu^* \in Q}\mu^*(s_1) \ge  p^2(s_1,\theta_1)=0.6,$$
which is impossible. \medskip

Example 4 shows that axiom ISU is generally violated under the max-min expected utility framework. This gives a justification for our framework where ambiguity only comes from signals, in which ISU assumption works.

\subsection{Comparison with the Proxy Rule}\label{subsec_proxy}
In this section, we compare CML with the proxy rule by \citet*{wp2018proxy}. To start with, we introduce the proxy rule. For simplicity, assume that $S \times \Theta$ is finite. Proxy rule works for totally monotone capacities. Totally monotone capacities have the following equivalent multi-prior characterization.  For any nonempty $E\subseteq S \times \Theta,$ let $\mathcal{P}_{E}$ be the set of all probability distributions that have support $E$, i.e., $\mathcal{P}_E=\{p \in \Delta(S\times\Theta): p(E)=1\}$. A totally monotone capacity has a multiple-prior representation given by 
\begin{equation}\label{eq_proxy}
\sum_{E \subseteq S\times \Theta: E \neq \emptyset}\alpha_{E} \mathcal{P}_{E},
\end{equation}
where $\alpha_{E} \in [0,1]$ for each nonempty $E \subseteq S\times \Theta$ and $\sum_{E \subseteq S\times \Theta: E\neq \emptyset} \alpha_E =1.$ Let $|E|$ denote the cardinality of event $E$. With the proxy rule, when $\theta$ is observed, the DM's set of posteriors is given by $$ \sum_{E \subseteq S\times\Theta: E \neq \emptyset}  \frac{\alpha_{E} \cdot |E\cap (S\times \theta)|\cdot |E|^{-1}}{\sum\limits_{E' \subseteq S\times\Theta: E' \neq \emptyset} \left( \alpha_{E'} \cdot |E'\cap (S\times \theta)|\cdot |E'|^{-1} \right) }  \mathcal{P}_{E\cap (S\times \theta)}.$$
A key property of the proxy rule is that ``not all news is bad news'': given an information structure and an extended act $f^*$, the DM's ex-post evaluation of $f^*|\theta$ should be weakly higher than his evaluation of $f^*$ under some signal $\theta$. The following example illustrates that CML violates ``not all news is bad news''. \medskip

\textbf{Example 5.} Let $S=\{s, s'\}$ and  $\Theta=\{\theta, \theta'\}$. The DM's prior set is $\mathcal{P}$, which is simple and consists of all convex combinations of $p^1,p^2,p^3$ and $p^4$, as shown in Table 4.

\begin{table}[H]
\caption*{\textbf{Table 4}}
\centering
\begin{tabular}{|c|c|c|}
\hline
  $p^1$   & $\theta$  & $\theta'$    \cr\hline
$s$ & $9/20$ & $1/20$     \cr\hline
$s'$ & $2/5$ & $1/10$     \cr\hline 
\end{tabular}
\quad
\begin{tabular}{|c|c|c|}
\hline
 $p^2$     & $\theta$  & $\theta'$    \cr\hline
$s$ & $1/20$ & $9/20$     \cr\hline
$s'$ & $1/10$ & $2/5$     \cr\hline
\end{tabular}

\bigskip

\begin{tabular}{|c|c|c|}
\hline
 $p^3$     & $\theta$  & $\theta'$    \cr\hline
$s$ & $9/20$ & $1/20$     \cr\hline
$s'$ & $1/10$ & $2/5$     \cr\hline
\end{tabular}
\quad
\begin{tabular}{|c|c|c|}
\hline
 $p^4$     & $\theta$  & $\theta'$    \cr\hline
$s$ & $1/20$ & $9/20$     \cr\hline
$s'$ & $2/5$ & $1/10$     \cr\hline
\end{tabular}
\end{table}

$\mathcal{P}$ admits a totally monotone capacity since it satisfies condition (\ref{eq_proxy}): $$\mathcal{P}= \frac{1}{20} \mathcal{P}_{\{(s,\theta)\}} + \frac{1}{20} \mathcal{P}_{\{(s,\theta')\}}+\frac{2}{5} \mathcal{P}_{\{(s,\theta), (s,\theta')\}} + \frac{1}{10}\mathcal{P}_{\{(s',\theta)\}}+ \frac{1}{10}\mathcal{P}_{\{(s',\theta')\}}+ \frac{3}{10}\mathcal{P}_{\{(s',\theta), (s',\theta')\}}.$$ Consider an extended act $f^*$: $f^*(s,\theta)=f^*(s,\theta')=0$ and $f^*(s',\theta)=f^*(s',\theta')=1$. The DM's prior over $S$ is $\mu=(1/2, 1/2)$ according to $\mathcal{P}$. When $\theta$ is realized, the DM's CML posterior over $S$ is $(\frac{9}{17}, \frac{8}{17})$. When $\theta'$ is realized, the DM's CML posterior is $(\frac{9}{17}, \frac{8}{17})$. Obviously, the DM lowers his evaluation of $f^*$ after each signal. As a result, CML violates ``not all news is bad news''. The next example shows that the proxy rule violates axiom ISU.\medskip

\textbf{Example 6.}  Let $S=\{s, s'\}$ and  $\Theta=\{\theta, \theta'\}$. The DM's prior set is  $\mathcal{P}$, which is simple and  consists of all convex combinations of $p^1$ and $p^2$.  $p^1$ and $p^2$ are shown in Table 5.

\begin{table}[H]
\caption*{\textbf{Table 5}}
\centering
\begin{tabular}{|c|c|c|}
\hline
  $p^1$   & $\theta$  & $\theta'$    \cr\hline
$s$ & $1/3$ & $1/6$     \cr\hline
$s'$ & $0$ & $1/2$     \cr\hline 
\end{tabular}
\quad
\begin{tabular}{|c|c|c|}
\hline
 $p^2$     & $\theta$  & $\theta'$    \cr\hline
$s$ & $1/3$ & $1/6$      \cr\hline
$s'$ & $1/2$ & $0$     \cr\hline
\end{tabular}
\end{table}

$\mathcal{P}$ admits a totally monotone capacity since $$\mathcal{P}= \frac{1}{3} \mathcal{P}_{\{(s,\theta)\}} + \frac{1}{6} \mathcal{P}_{\{(s,\theta')\}} + \frac{1}{2}\mathcal{P}_{\{(s',\theta), (s',\theta')\}}.$$ The DM's prior over $\{s,s'\}$ is $\mu=(1/2, 1/2)$ according to $\mathcal{P}$. With the proxy rule, when $\theta$ is observed, the DM's posterior over $S$ is $\mu^*=(4/7,3/7)$. Consider an extended act $f^*:$ $f^*(s,\theta)=f(s',\theta')=1,$ $f^*(s',\theta)=0$ and $f^*(s,\theta')=10/7.$ We can verify that the evaluation $\mathbb{E}_{\mathcal{P}}(f^*)$ is $4/7$. With the proxy rule, the ex-post evaluation of $f^*|\theta$ is given by $\mathbb{E}_{\mu^*}(f^*|\theta)$, which is again equal to $4/7.$ Consider another extended act $g^*$ where $g^*(s,\theta)=g(s',\theta')=1,$ $g^*(s',\theta)=1/2$ and $g^*(s,\theta')=10/7.$ $g^*$ differs from $f^*$ only at $(s',\theta)$, where $g^*$ yields a higher payoff. We have $\mathbb{E}_{\mathcal{P}}(g^*)=23/28 > \mathbb{E}_{\mu^*}(g^*|\theta)=11/14$.  Hence, axiom ISU is violated by the proxy rule.

\section{Conclusion}\label{sec_conclusion}
In this paper, we axiomatize a new updating rule, CML,  for updating ambiguous information. Different from existing rules, CML satisfies and can be characterized by axiom ISU. We show that CML satisfies divisibility, accommodates recent experimental findings and has simple predictions on learning. When an agent updates according to CML, we show that an information designer can benefit from introducing ambiguous information and almost achieves the maximal payoff. \medskip

We propose two streams of future works. First, axiom ISU can be tested in the lab. Testing axiom ISU is straightforward once we collect DMs' ex-ante and ex-post evaluation functions.  Second, we can investigate whether axiom ISU is compatible with other theoretical frameworks of ambiguity, e.g., the dual-self expected utility framework.\footnote{See \citet*{wp2020boolean} for the dual-self expected utility theory.}

\newpage

\section{Appendix}
\subsection{Appendix A: Proofs of Propositions and Theorems} 
\begin{proof}[Proof of Proposition \ref{prop_isau}]
Let $V \in \mathcal{V}$ be represented by $\mathcal{P}$. Since $\mathcal{P}$ is simple, we can write $\mathcal{P}$ as  $(\mu, \{c^t(\cdot|s)_{s \in S}\}_{t \in T})$. 
Consider $f^*$ and $g^*$ satisfying the conditions stated in the proposition. We have 
\begin{align*}
V(g^*)-V(f^*)  = & \mathbb{E}_{\mathcal{P}} (g^*) - \mathbb{E}_{\mathcal{P}} (f^*) \\
= & \mathbb{E}_{\mathcal{P}} (g^*) - \mathbb{E}_{p^*} (f^*) \\
\le & \mathbb{E}_{p^*} (g^*) - \mathbb{E}_{p^*} (f^*) \\
= & p^*(s,\theta) (g^*(s,\theta)-f^*(s,\theta)) \\
\le & \max_{p \in \mathcal{P}} p(s,\theta) (g^*(s,\theta)-f^*(s,\theta)) \\
= & \mu(s)\max_{t \in \mathcal{T}} c^t(\theta|s) (g^*(s,\theta)-f^*(s,\theta)),
\end{align*}
where $p^* \in \mathcal{P}$ minimizes the expectation of $f^*$. Other inequalities are obvious. Let $\mu_{\theta}$ be the CML posterior when signal $\theta$ is observed. For $V_{\theta}(g^*|\theta)$ and $V_{\theta}(f^*|\theta)$, we have \begin{align*}
 V_{\theta}(g^*|\theta) - V_{\theta}(f^*|\theta)  = & \mu_{\theta}(s)(g^*(s,\theta)-f^*(s,\theta)) \\
= &  \frac{\mu(s) \max_{t \in \mathcal{T}} c^t(\theta|s) }{ \sum_{s' \in S} \mu(s') \max_{t \in \mathcal{T}} c^t(\theta|s')} (g^*(s,\theta)-f^*(s,\theta))  \\
\ge & \mu(s) \max_{t \in \mathcal{T}} c^t(\theta|s) (g^*(s,\theta)-f^*(s,\theta)).
\end{align*}
The last inequality holds since $\sum_{s' \in S} \mu(s') \max_{t \in \mathcal{T}} c^t(\theta|s') \le \sum_{s' \in S} \mu(s')= 1.$ The proposition is thus shown. 
\end{proof}\bigskip

\begin{proof}[Proof of Theorem \ref{thm_main}]
Let $V \in \mathcal{V}$ be represented by $\mathcal{P}$. It is obvious that $\theta \in \Theta_V$
if and only if there exists $p \in \mathcal{P}$ such that $p(s,\theta)>0$ for some $s$.  For any nonempty, convex and closed set of probability distributions $\mathcal{Q} \subseteq \Delta(S)$ and any $S' \subseteq S$, if $\min_{\mu \in \mathcal{Q}} \mu(S') > 0$, let $\mathcal{Q}|S'$ be the set of conditional probabilities of $\mathcal{Q}$ on $S'$. That is, $\mu_{S'} \in \mathcal{Q}|S'$ if and only if $\mu_{S'}$ is the Bayesian posterior of some $\mu \in \mathcal{Q}$ on $S'$. Since $\min_{\mu \in \mathcal{Q}} \mu(S') > 0$, $\mathcal{Q}|S'$ is nonempty, convex and closed. For any evaluation function $U$ that is represented by $\mathcal{Q}$, $U$ is strictly increasing on $S' \subseteq S$ if and only if $\min_{\mu \in \mathcal{Q}} \mu(S') > 0$. Before proceeding, we first prove a lemma.
\begin{lemma}\label{lm_ratio_consistency}
For any $\{s, s'\} \subseteq S$ and any two evaluation functions $U$ and $U'$, represented by $\mathcal{Q}$ and $\mathcal{Q}'$ respectively, if $\min_{\mu \in \mathcal{Q}} \mu(\{s,s'\}) > 0$ and $\min_{\mu' \in \mathcal{Q}'} \mu'(\{s_1,s_2\})>0$, then the following two conditions are equivalent. 
\begin{enumerate}
\item For any $x,y,z \in \mathbb{K},$ $U(x[s]y[s']z)=z$ if and only if $U'(x[s]y[s']z)=z$.
\item  $\mathcal{Q}|\{s,s'\}=\mathcal{Q}'|\{s,s'\}.$
\end{enumerate} 
\end{lemma}

\begin{proof}[Proof of Lemma \ref{lm_ratio_consistency}] Assume  that condition 1 holds. We prove condition 2. Assume to the contrary that $\mathcal{Q}|\{s, s'\} \neq \mathcal{Q}'|\{s, s'\}$. There exist $x, y \in \mathbb{K}$ such that $$ \min_{\mu_{\{s,s'\}} \in \mathcal{Q}|\{s, s'\}}  \left( \mu_{\{s,s'\}}(s)x+ \mu_{\{s,s'\}}(s')y \right) = z > \min_{\mu'_{\{s,s'\}} \in \mathcal{Q}'|\{s, s'\}} \left( \mu'_{\{s,s'\}}(s)x+ \mu'_{\{s,s'\}}(s')y \right).$$
This implies that $U(x[s]y[s']z)=z$ and $U'(x[s]y[s']z)<z$, which is a contradiction. \medskip

Inversely, assume that condition 2 holds. We only need to show that \begin{equation}\label{eq_lemma_1}
U(x[s]y[s']z)=z
\end{equation} if and only if 
\begin{equation}\label{eq_lemma_2}
\min_{\mu_{\{s,s'\}} \in \mathcal{Q}|\{s,s'\}} \left( \mu_{\{s,s'\}}(s)x+\mu_{\{s,s'\}}(s')y \right) = z.
\end{equation}
Then, condition 2 implies condition 1.\medskip

Suppose that condition (\ref{eq_lemma_1}) holds. There exists $\mu^* \in \mathcal{Q}$ such that $\mu^*(s) x +\mu^*(s')y + (1-\mu^*(\{s,s'\}))z=z$. Thus, $\mu^*(s) x +\mu^*(s')y=(\mu^*(s)    +\mu^*(s')) z$. Since $\min_{\mu \in \mathcal{Q}} \mu(\{s,s'\}) > 0$, we have $\mu^*(s)+\mu^*(s')>0$ and $$\frac{\mu^*(s)}{\mu^*(s)+\mu^*(s')} x + \frac{\mu^*(s')}{\mu^*(s)+\mu^*(s')} y=z.$$ Note that $\mu^*_{\{s,s'\}}=(\frac{\mu^*(s)}{\mu^*(s)+\mu^*(s')}, \frac{\mu^*(s')}{\mu^*(s)+\mu^*(s')}) \in \mathcal{Q}|\{s,s'\}$. Hence,  we have $$\min_{\mu_{\{s,s'\}} \in \mathcal{Q}|\{s,s'\}} \left( \mu_{\{s,s'\}}(s)x+\mu_{\{s,s'\}}(s')y \right) \le z.$$ If the above inequality holds strictly, by the assumption that $\min_{\mu \in \mathcal{Q}} \mu(\{s,s'\}) > 0$, we know $U(x[s]y[s']z)<z$, which is a contradiction. Hence, condition (\ref{eq_lemma_2}) must hold. Showing that condition (\ref{eq_lemma_2}) implies condition (\ref{eq_lemma_1}) is similar. 
\end{proof} \medskip

\textbf{(Necessity.)} For axiom IIS, consider $V,W \in \mathcal{V}$. Let $V$ be represented by $\mathcal{P}$ and $W$ represented by $\mathcal{P}'$. If $V(f^*[S\times\theta]x)=W(f^*[S\times\theta]x)$ for all $f^* \in \mathcal{F}^*$ and $x \in \mathbb{K},$ then we have $\mathcal{P}\approx^{S\times\theta} \mathcal{P}'$. Therefore, $\theta \in \Theta_V$ implies that $\theta \in \Theta_W.$ Moreover, for each $s \in S,$ $\max_{p \in \mathcal{P}} p(s,\theta) = \max_{p' \in \mathcal{P}'} p'(s,\theta).$ By the formula of CML posterior (\ref{eq_d1}), we have $V_{\theta}= W_{\theta}.$ \medskip

For axiom RC, consider $V,W \in \mathcal{V}$. Let $V$ be represented by $\mathcal{P}$ and $W$ represented by $\mathcal{P}'$. If $V(f^*[\{s,s'\}\times \Theta]x)=W(f^*[\{s,s'\}\times \Theta]x)$ for all $f^* \in \mathcal{F}^*$ and $x \in \mathbb{K}$, then $\mathcal{P} \approx^{\{s,s'\} \times \Theta} \mathcal{P}'.$ For any $\theta \in \Theta_V \cap \Theta_W,$ we know 
\begin{equation}\label{eq_necessary_1}
\begin{split}
\max_{p \in \mathcal{P}}p(s,\theta) = \max_{p' \in \mathcal{P}'}p'(s,\theta), \text{ and } \max_{p \in \mathcal{P}}p(s',\theta) = \max_{p' \in \mathcal{P}'}p'(s',\theta).\\
\end{split}
\end{equation}
Let $V_{\theta}$ be represented by $\mathcal{Q}_{\theta}=\{\mu_{\theta}\}$ and $W_{\theta}$ represented by $\mathcal{Q}'_{\theta}=\{\mu'_{\theta}\}$. Since $V_{\theta}$ and $W_{\theta}$ are strictly increasing on $\{s,s'\}$, we have $\mu_{\theta}(\{s,s'\})>0$ and $\mu'_{\theta}(\{s,s'\})>0$. By the CML posterior formula (\ref{eq_d1}) and condition (\ref{eq_necessary_1}), we have  $\mathcal{Q}_{\theta}|\{s,s'\}=\mathcal{Q}'_{\theta}|\{s,s'\}$. By Lemma \ref{lm_ratio_consistency}, axiom RC holds. Axiom ISU is shown by Proposition \ref{prop_isau}.

\bigskip

\textbf{(Sufficiency.)} Assume that axioms IIS,RC and ISU all hold. Through out the proof of the sufficiency part, assume  the payoff space to be $\mathbb{R}$. This is without loss of generality since for any payoff $y$ not in $\mathbb{K}$, we can pick $x$ in the interior of $\mathbb{K}$ and take the convex combination  $\alpha x + (1-\alpha)y$ of $x$ and $y$ such that $\alpha x + (1-\alpha)y$ is in $\mathbb{K}$. \medskip

\begin{lemma}\label{lm_full}
Let $V \in \mathcal{V}$ be represented by $\mathcal{P}=(\mu, \{c^t(\cdot|s)_{s \in S}\}_{t \in T})$ and $V_{\theta}$ represented by $\mathcal{Q}_{\theta}$, where $\theta \in \Theta_{V}.$ If \begin{equation}\label{eq_full}
\sum_{s \in S} \mu(s)\max_{t \in T}c^t(\theta|s) =1,
\end{equation}
then $\mathcal{Q}_{\theta}=\{\mu\}$.
\end{lemma}

\begin{proof}
Fix some $s \in S$ with $\mu(s)>0$. By condition (\ref{eq_full}), $\max_{t \in T}c^t(\theta|s)=1.$ Consider an extended act $f^*$ satisfying that 
\begin{align*}
& f^*(s,\theta)=0,\\
& f^*(s,\theta')=1, \forall \theta' \in \Theta \backslash \theta, \\
& f^*(s',\theta')=0, \forall (s',\theta') \in (S\times\Theta)\backslash (s\times\Theta). 
\end{align*}
For such an extended act, we have $$V(f^*)= \mu(s)(1-\max_{t \in T}c^t(\theta|s)) = 0 = V_{\theta}(f^*|\theta).$$
Consider an extended act $g^*$ satisfying that
\begin{align*}
& g^*(s,\theta)=\frac{1}{2},\\
& g^*(s,\theta')=1, \forall \theta' \in \Theta \backslash \theta, \\
& g^*(s',\theta')=0, \forall (s',\theta') \in (S\times\Theta)\backslash (s\times\Theta).
\end{align*}  
We have $$V(g^*)=\frac{1}{2} \mu(s),\text{ and } V_{\theta}(g^*|\theta)=\frac{1}{2}\min_{\mu^* \in \mathcal{Q}_{\theta}} \mu^*(s).$$
For extended acts $f^*$ and $g^*$, axiom ISU implies that 
$$\min_{\mu^* \in \mathcal{Q}_{\theta}} \mu^*(s) \ge \mu(s).$$
Since this holds for each $s \in S$ with $\mu(s)>0$, we  have for each $\mu^* \in \mathcal{Q}_{\theta}$ and each $s \in S$, $\mu^*(s)=\mu(s).$ That is, $\mathcal{Q}_{\theta}=\{\mu\}$.
\end{proof}

\begin{lemma}\label{lm_final}
Let $V \in \mathcal{V}$ be represented by $\mathcal{P}=(\mu, \{c^t(\cdot|s)_{s \in S}\}_{t \in T})$ and $V_{\theta}$   represented by $\mathcal{Q}_{\theta},$ where $\theta \in \Theta_{V}.$ If $\mu(s_1)\max_{t \in T}c^t(\theta|s_1)>0$, then for any $s_2 \in S$ that is different from $s_1$ and any $\mu^* \in \mathcal{Q}_{\theta}$,  it holds that $\mu^*(s_1)>0$ and
$$\frac{\mu^*(s_2)}{\mu^*(s_1)} = \frac{\mu(s_2)\max_{t \in T}c^t(\theta|s_2)}{\mu(s_1)\max_{t \in T}c^t(\theta|s_1)}.$$
\end{lemma}

\begin{proof}
Fix $s_1$ and $s_2$ that satisfy the conditions stated in the lemma. Pick $s_3$ from $S\backslash \{s_1, s_2\}.$ Consider $V' \in \mathcal{V}$, represented by $\mathcal{P}'=(\mu', \{b^t(\cdot|s)_{s \in S}\}_{t \in T})$, where $(\mu', \{b^t(\cdot|s)_{s \in S}\}_{t \in T})$ satisfies the following conditions:\medskip

(1) $\mu'(s_1)=\mu(s_1)\max_{t \in T}c^t(\theta|s_1)$ and $\mu'(s_2)=\mu(s_2)\max_{t \in T}c^t(\theta|s_2)$.

(2) $\mu'(s_3)=\mu(s_1)+\mu(s_2)+\mu(s_3)-\mu'(s_1)-\mu'(s_2)$.

(3) $\mu'(\hat{s})=\mu(\hat{s})$ for each $\hat{s} \in S\backslash \{s_1,s_2,s_3\}$.

(4) $b^t(\theta|s_1)=c^t(\theta|s_1) \left( \max_{\hat{t} \in T}c^{\hat{t}}(\theta|s_1) \right)^{-1}$ for all $t \in T.$ 

(5) $b^t(\theta|s_2)=c^t(\theta|s_2) \left( \max_{\hat{t} \in T}c^{\hat{t}}(\theta|s_2) \right)^{-1}$ for all $t \in T$ if $\mu'(s_2)> 0$.

(6) $b^t(\theta|s_3)=0$ for all $t \in T$ if $\mu(s_3)=0$.

(7) $b^t(\theta|s_3)=\frac{\mu(s_3)}{\mu'(s_3)} c^t(\theta|s_3)$ for all $t \in T$ if $\mu(s_3)>0.$

(8) $b^t(\theta|\hat{s})=c^t(\theta|\hat{s})$ for all $\hat{s} \in S\backslash \{s_1,s_2,s_3\}$ and all $t \in T$. \medskip

With the above conditions, we can verify that for each $t \in T$ and each $s \in S$, $\mu(s)c^t(\theta|s)= \mu'(s)b^t(\theta|s)$. Therefore, $\mathcal{P} \approx^{S\times\theta} \mathcal{P}'$. By axiom IIS, we know $V_{\theta} = V'_{\theta}$. Let $V'_\theta$ be represented by $\mathcal{Q}'_\theta$. We have $\mathcal{Q}_\theta= \mathcal{Q}'_\theta$. By a similar argument as in Lemma \ref{lm_full}, we can show that $\mu^*(s_1) \ge \mu'(s_1) > 0$ for all $\mu^* \in \mathcal{Q}'_\theta > 0$.  \medskip

Consider $V'' \in \mathcal{V}$, represented by $\mathcal{P}''=(\mu'', \{a^t(\cdot|s)_{s \in S}\}_{t \in T})$, where $(\mu'', \{a^t(\cdot|s)_{s \in S}\}_{t \in T})$ satisfies the following conditions:\medskip

(1) $\mu''(s)=\mu'(s)$ for all $s \in S$.

(2) $a^t(\theta'|s_1)=b^t(\theta'|s_1)$ and $a^t(\theta'|s_2)=b^t(\theta'|s_2)$ for all $\theta' \in \Theta$ and all $t \in T$.

(3) $a^t(\theta|\hat{s}) = 1$ for all $\hat{s} \in S\backslash \{s_1,s_2\}$ and all $t \in T$. \medskip

Let $V''_{\theta}$ be represented by $\mathcal{Q}''_{\theta}.$ With the above three conditions, we have $$\sum_{s \in S} \mu''(s)\max_{t \in T} a^t(\theta|s)=1.$$ 
By Lemma \ref{lm_full}, we have $\mathcal{Q}''_{\theta}=\{\mu''\}=\{\mu'\}.$ We can verify that $\mu'(s)b^t(\theta'|s)=\mu''(s) a^t(\theta'|s)$ for each $s \in \{s_1, s_2\}$, each $\theta' \in \Theta$ and each $t \in T$.  Hence, we have $\mathcal{P}' \approx^{\{s_1,s_2\}\times\Theta} \mathcal{P}''.$ Since  $\mu^*(s_1) > 0$ for each $\mu^* \in \mathcal{Q}'_\theta$ and $\mathcal{Q}''_\theta=\{\mu'\}$ satisfies $\mu'(s_1)>0$, $V'_\theta$ and $V''_\theta$ are both strictly increasing on $\{s_1, s_2\}$. By axiom RC and Lemma \ref{lm_ratio_consistency}, we know that for each $\mu^* \in \mathcal{Q}'_{\theta},$  $$\frac{\mu^*(s_2)}{\mu^*(s_1)} =  \frac{\mu''(s_2)}{\mu''(s_1)} = \frac{\mu'(s_2)}{\mu'(s_1)}= \frac{\mu(s_2)\max_{t\in T}c^t(\theta|s_2)}{\mu(s_1)\max_{t\in T}c^t(\theta|s_1)}.$$
Since $\mathcal{Q}_\theta=\mathcal{Q}'_\theta$, for each $\mu^* \in \mathcal{Q}_\theta$, the above condition holds. This finishes the proof of the lemma.
\end{proof}

By Lemma \ref{lm_final}, if $\mu(\hat{s})\max_{t\in T}c^t(\theta|\hat{s})>0$ for some $\hat{s} \in S$, then for each $\mu^* \in \mathcal{Q}_\theta$, $\mu^*(\hat{s})>0$  and $$\frac{\mu^*(s)}{\mu^*(\hat{s})} =  \frac{\mu(s)\max_{t\in T}c^t(\theta|s)}{\mu(\hat{s})\max_{t\in T}c^t(\theta|\hat{s})}$$ for each $s \in S.$ Hence, for each $\mu^* \in \mathcal{Q}_\theta$ and each $s \in S$, $$\mu^*(s)= \frac{\mu^*(s)}{\sum_{s' \in S} \mu^*(s')} =\frac{\frac{\mu^*(s)}{\mu^*(\hat{s})}}{\sum_{s' \in S} \frac{\mu^*(s')}{\mu^*(\hat{s})}} = \frac{\mu(s)\max_{t \in T}c^t(\theta|s)}{\sum_{s' \in S}\mu(s')\max_{t \in T}c^t(\theta|s')}.$$ We are done.
\end{proof}\bigskip

\begin{proof}[Proof of Theorem \ref{thm_rational}]
For necessity, suppose that $\{(c^t(\cdot|s))_{s\in S}\}_{t \in T}$ is the DM's set of interpretations of signals that satisfies conditions (\ref{eq_nonnull}) and (\ref{eq_cml_rational}). For each $s \in S$ and each $\theta \in \Theta^*$, 
\begin{align*}
\frac{\mu_{\theta}(s)}{\mu(s)}   \left( \max_{s' \in S'}\frac{\mu_{\theta}(s')}{\mu(s')} \right)^{-1} & = \frac{\max_{t \in T}c^t(\theta|s)}{\max_{s' \in S} \left(  \max_{t \in T}c^t(\theta|s') \right) }  \ge \max_{t \in T} c^t(\theta|s).
\end{align*}
Therefore, $$ \sum_{\theta \in \Theta^*} \frac{\mu_{\theta}(s)}{\mu(s)}   \left( \max_{s' \in S'}\frac{\mu_{\theta}(s')}{\mu(s')} \right)^{-1} \ge \sum_{\theta \in \Theta^*} \left(  \max_{t \in T} c^t(\theta|s) \right)  \ge 1.$$ This shows the necessity part. \medskip

For sufficiency, suppose that the belief profile $(\mu, \{\mu_{\theta}\}_{\theta \in \Theta^*})$ satisfies that for each $s\in S,$ \begin{equation}\label{eq_ge1}
 \sum_{\theta \in \Theta^*} \frac{\mu_{\theta}(s)}{\mu(s)}   \left( \max_{s' \in S'}\frac{\mu_{\theta}(s')}{\mu(s')} \right)^{-1} \ge 1.
\end{equation}
We construct the finite set of interpretations. Define $\{c^{\theta}(\cdot|s)_{s \in S}\}_{\theta \in \Theta^*},$ where each interpretation $c^{\theta}$ is labeled by a signal $\theta\in \Theta^*.$  Let $S_{\theta} := \arg\max_{s \in S} \left( \mu_{\theta}(s)/\mu(s) \right).$ For each $s \in S$ and $\theta \in \Theta^*$, let $$c^{\theta}(\theta|s)= \frac{\mu_{\theta}(s)}{\mu(s)} \left( \max_{s' \in S'}\frac{\mu_{\theta}(s')}{\mu(s')} \right)^{-1}.$$ Obviously, $c^{\theta}(\theta|s) \le 1$ for each $s \in S$ and $c^{\theta}(\theta|s) = 1$ for each $s \in S_\theta.$ By condition (\ref{eq_ge1}), we have $$\sum_{\theta \in \Theta^*}c^{\theta}(\theta|s) \ge 1$$ for each $s \in S$. Hence, we can find non-negative numbers $\{c^{\theta}(\theta'|s)\}_{\theta' \in \Theta^* \backslash \theta}$ for each $\theta \in \Theta^*$ and each $s \in S$ such that $$ \sum_{\theta' \in \Theta^*} c^{\theta}(\theta'|s)= 1,$$ and
$$ c^{\theta}(\theta'|s) \le  c^{\theta'}(\theta'|s).$$ 
Since $\{c^{\theta'}(\theta|s)\}_{\theta' \in \Theta^*}$ is maximized at $\theta'=\theta$ for each $s$ and each $\theta$, we have the desired posterior $\mu_{\theta}$ for each $\theta \in \Theta^*.$ 
\end{proof}\bigskip

\begin{proof}[Proof of Theorem \ref{thm_indesign}]
If there exists $a^* \in \bigcap_{s \in S} \arg\max_{a \in A^*} v(s,a)$, let $a^s=a^*$ for each $s \in S.$ Otherwise, for each $s \in S,$ pick $a^s \in \arg\max_{a \in A^*} v(s,a).$ Pick $\mu^s \in \Delta^o(S)$ such that $a^s \in \mathcal{BR}(\mu^s).$ $\{\mu^s\}_{s \in S}$ are the target posteriors. For each $\mu^s,$ pick a vector $(\lambda_{s'}^{s})_{s' \in S} \in (0,1)^{S}$ such that 
$$\frac{\mu(s') \lambda_{s'}^{s}}{\mu(s'') \lambda_{s''}^{s}} = \frac{\mu^s(s')}{\mu^s(s'')}$$
for each $s',s'' \in S.$ 
Fix the vector $(\lambda_{s'}^{s})_{s'\in S}$ for each $s \in S.$ For any $\{r^s\}_{s \in S} \subseteq (0,1),$ let $n(\{r^s\}_{s \in S})$ be an integer satisfying that  \begin{equation}\label{eq_implement_0}
\max_{s'\in S, s'' \in S}\frac{1-r^{s'}\lambda_{s''}^{s'}}{n(\{r^s\}_{s \in S})-1} < \min_{s' \in S, s'' \in S} r^{s'}\lambda_{s''}^{s'}.
\end{equation}
For the given $\epsilon >0$, pick $\{r^s\}_{s \in S} \subseteq (0,1)$ such that  
\begin{equation}\label{eq_implement_1}
\forall s \in S, \sum_{s' \in S} \mu(s')(r^{s}\lambda_{s'}^{s}v(s',a^{s}) + (1-r^{s}\lambda_{s'}^{s})v(s',a^{s'})) = l^*,
\end{equation}
where $l^*$ satisfies that
\begin{equation}\label{eq_implement_2}
l^* > \sum_{s \in S} \left( \mu(s)\max_{a \in A^*} v(s,a) \right) -\epsilon.
\end{equation}
In the case $a^s=a^*,$ we have
$$ l^* = \sum_{s \in S} \left( \mu(s)\max_{a \in A^*} v(s,a) \right),$$ which implies that $\{r^s\}_{s \in S}$ exist.
For the case where $\bigcap_{s \in S} \arg\max_{a \in A^*} v(s,a) = \emptyset$, since as $r^s$ converges to zero, $$\sum_{s' \in S} \mu(s')(r^{s}\lambda_{s'}^{s}v(s',a^{s}) + (1-r^{s}\lambda_{s'}^{s})v(s',a^{s'}))$$ is strictly lower than but converges to $$ \sum_{s' \in S} \mu(s')v(s',a^{s'})= \sum_{s \in S} \left( \mu(s)\max_{a \in A^*} v(s,a) \right), $$ 
the numbers $\{r^s\}_{s \in S}$ that satisfy the desired conditions exist. Fix the numbers $\{r^s\}_{s \in S}$.\medskip

We proceed to construct the information structure. Let $N=n(\{r^s\}_{s \in S})$. Let $S=\{s_1,...,s_m\}.$ Let $\Theta=\{\theta_{j,l}\}_{j \in \{1,...,m\}, l\in \{1,...,N\}}.$ Let there be $m\cdot N$ signal generating systems. Each signal generating system is denoted by $c^{j,l}$ for some $j \in \{1,...,m\}$ and $l \in \{1,...,N\}.$ For each $c^{j,l}$, let 
\begin{align*}
& c^{j,l}(\theta_{j,l}|s)= r^{s_j}\lambda_{s}^{s_j}, \forall s \in S, \\
& c^{j,l}(\theta_{j',l'}|s)= \frac{1-r^{s_j}\lambda_{s}^{s_{j}}}{N -1}, \text{ if } s=s_{j'} \text{ and }  l \neq l', \text{ and} \\
& c^{j,l}(\theta_{j',l'}|s)=0 \text{ otherwise.}
\end{align*} 
By condition (\ref{eq_implement_0}), for each $\theta_{j,l},$ its conditional maximum probability on each $s \in S$ is achieved uniquely by $c^{j,l}$ and equal to $r^{s_j}\lambda^{s_j}_{s}$. Hence, the CML posterior at signal $\theta_{j,l}$ is  $\mu^{s_j}$. \medskip

Consider $G= (\Theta, \{c^{j,l}(\cdot|s)_{s \in S}\}_{j \in \{1,...,m\}, l\in\{1,...,N\}}, \{a^{\theta_{j,l}}\}_{\theta_{j,l} \in \Theta}),$ where $a^{\theta_{j,l}}=a^{s_j}.$ We argue $G$ satisfies the desired conditions of the theorem. Note that $a^{\theta_{j,l}}$ is indeed the optimal action of the agent given the CML posterior $\mu^{s_j}$. For each signal generating system $c^{j,l},$ conditional on state $s_j$, action $a^{s_j}$ is taken by the agent with probability $1$. Conditional on state $s \neq s_j$, action $a^{s}$ is taken by the agent with probability $1-\frac{1-r^{s_j}\lambda_{s}^{s_j}}{N -1},$ and action $a^{s_j}$ is taken with probability $\frac{1-r^{s_j}\lambda_{s}^{s_j}}{N -1}.$ By conditions (\ref{eq_implement_1}) and (\ref{eq_implement_2}), we know that $G$ satisfies the conditions in the statement of the theorem.
\end{proof}

\subsection{Appendix B: Other Proofs}\label{subsec_appendix_b}
In this section, we show that any two of the three axioms are not sufficient for CML.  \medskip

\textbf{Axiom IIS and Axiom RC.} FB satisfies both axiom IIS and axiom RC and violates axiom ISU. Let $V \in \mathcal{V}$ be represented by $\mathcal{P}$. If $\theta \in \Theta_V$, the ex-post evaluation $V_{\theta}$ specified by FB is represented by the closure of $\mathcal{P}|\theta$, denoted by $cl(\mathcal{P}|\theta)$. Obviously, FB satisfies axiom IIS. \medskip

We show that FB satisfies axiom RC. Let $V \in \mathcal{V}$ and $W \in \mathcal{V}$ be represented by $\mathcal{P}$ and $\mathcal{P}'$ respectively. Let $\theta \in \Theta_V \cap \Theta_W$.
With FB, the two evaluation functions $V_{\theta}$ and $W_{\theta}$ are represented by $cl(\mathcal{P}|\theta)$ and $cl(\mathcal{P}'|\theta)$ respectively. Fix two distinct states $s$ and $s'$. To show axiom RC, we need to show that $\mathcal{P} \approx^{\{s,s'\} \times \Theta} \mathcal{P}'$, $\inf_{\mu \in \mathcal{P}|\theta}(\{s,s'\}) > 0$ and $\inf_{\mu' \in \mathcal{P}'|\theta}(\{s,s'\}) > 0$ implies $cl(\mathcal{P}|\theta)|\{s,s'\} = cl(\mathcal{P}'|\theta)|\{s,s'\}$. Then, by Lemma \ref{lm_ratio_consistency}, we are done. Note that $$\min_{\mu_{\{s,s'\}} \in cl(\mathcal{P}|\theta)|\{s,s'\}} \frac{\mu_{\{s,s'\}}(s)}{\mu_{\{s,s'\}}(s')} = \inf_{p \in \mathcal{P}: p(s,\theta)+p(s',\theta)>0} \frac{p(s,\theta)}{p(s',\theta)}, $$
and 
$$\min_{\mu'_{\{s,s'\}} \in cl(\mathcal{P}'|\theta)|\{s,s'\}} \frac{\mu'_{\{s,s'\}}(s)}{\mu'_{\{s,s'\}}(s')} = \inf_{p' \in \mathcal{P}': p'(s,\theta)+p'(s',\theta)>0} \frac{p'(s,\theta)}{p'(s',\theta)}.$$
Since $\mathcal{P} \approx^{\{s,s'\} \times \Theta} \mathcal{P}'$, we have $$\min_{\mu_{\{s,s'\}} \in cl(\mathcal{P}|\theta)|\{s,s'\}} \frac{\mu_{\{s,s'\}}(s)}{\mu_{\{s,s'\}}(s')}  = \min_{\mu'_{\{s,s'\}} \in cl(\mathcal{P}'|\theta)|\{s,s'\}} \frac{\mu'_{\{s,s'\}}(s)}{\mu'_{\{s,s'\}}(s')}.$$ 
Similarly, we have $$\max_{\mu_{\{s,s'\}} \in cl(\mathcal{P}|\theta)|\{s,s'\}} \frac{\mu_{\{s,s'\}}(s)}{\mu_{\{s,s'\}}(s')}  = \max_{\mu'_{\{s,s'\}} \in cl(\mathcal{P}'|\theta)|\{s,s'\}} \frac{\mu'_{\{s,s'\}}(s)}{\mu'_{\{s,s'\}}(s')}.$$
This implies that $cl(\mathcal{P}|\theta)|\{s,s'\} = cl(\mathcal{P}'|\theta)|\{s,s'\}$. We are done.\bigskip

\textbf{Axiom IIS and Axiom ISU.}
Consider the following updating rule. For any $V \in \mathcal{V}$ that is represented by $\mathcal{P}$ and any $\theta \in \Theta_V$, the evaluation function $V_{\theta}$ is represented by $$\mathcal{Q}_{\theta}=\left( \sum_{s \in S}  \max_{p \in \mathcal{P}}p(s,\theta) \right)\mu_{\theta} + \left( 1- \left( \sum_{s \in S}  \max_{p \in \mathcal{P}}p(s,\theta) \right) \right)cl(\mathcal{P}|\theta),$$
where $\mu_{\theta}$ is the CML posterior. That is, the DM's posterior set is a convex combination of the CML posterior set and the FB posterior set, where weights are given by $$\sum_{s \in S}  \max_{p \in \mathcal{P}}p(s,\theta) \text{ and } 1-\left( \sum_{s \in S}  \max_{p \in \mathcal{P}}p(s,\theta) \right).$$  
This updating rule obviously satisfies axiom IIS. For axiom ISU, note that for any $\mu^* \in \mathcal{Q}_{\theta},$ $$\mu^*(s) \ge \left( \sum_{s \in S}  \max_{p \in \mathcal{P}}p(s,\theta) \right)\mu_{\theta}(s) = \max_{p \in \mathcal{P}} p(s,\theta).$$ This indicates that the ex-post sensitivity is at least $\max_{p \in \mathcal{P}} p(s,\theta)$. Therefore, axiom ISU is satisfied.
\bigskip

\textbf{Axiom RC and Axiom ISU.}
Consider the following updating rule. For any $V \in \mathcal{V}$ that is represented by $\mathcal{P}=(\mu, \{c^t(\cdot|s)_{s \in S}\}_{t \in T})$ and any $\theta \in \Theta_V$, the evaluation function $V_{\theta}$ is represented by $\mu^{\lambda}_{\theta}$ for some $\lambda \in (0,1)$ where $$ \mu^{\lambda}_{\theta}(s) = \frac{\mu(s)\left( \max_{t \in T}c^t(\theta|s)\right)^{\lambda}}{\sum_{s' \in S}\mu(s')\left( \max_{t \in T}c^t(\theta|s')\right)^{\lambda}}, \forall s \in S.$$
Checking that this updating rule satisfies axiom RC is the same as checking that CML satisfies axiom RC. This rule also satisfies axiom ISU since $$\mu_{\theta}^\lambda(s) \ge \mu(s) \left( \max_{t \in T}c^t(\theta|s)\right)^{\lambda} \ge \mu(s) \max_{t \in T}c^t(\theta|s).$$

\newpage
\bibliographystyle{ecta} 
\bibliography{cml}
\end{document}